%% file: hierarchy.tex
\documentclass[aps, prb, twocolumn, superscriptaddress, floatfix, letterpaper, nofootinbib ]{revtex4-2}

\input{defs}

\usepackage{xcolor}

\usepackage{hyperref}
\hypersetup{
    colorlinks = true,
    allcolors = blue,
}

\begin{document}

\title{Hierarchy construction for non-abelian fractional quantum Hall states\\ 
via anyon condensation}

\author{Carolyn Zhang}
\affiliation{Department of Physics, Harvard University, Cambridge, Massachusetts 02138, USA}

\author{Ashvin Vishwanath}
\affiliation{Department of Physics, Harvard University, Cambridge, Massachusetts 02138, USA}

\author{Xiao-Gang Wen}
\affiliation{Department of Physics, Massachusetts Institute of Technology,
Cambridge, Massachusetts 02139, USA}

\begin{abstract} 
For a given parent fractional quantum Hall (FQH) state at filling fraction $\nu$, the hierarchy construction produces FQH states at nearby filling fractions $\{\nu_n\}$ by condensing minimally charged quasiholes or quasiparticles of the parent state into their own FQH states. The hierarchy construction has been useful for relating families of FQH states and for the experimental identification of the topological order of  parent states via the properties of daughter states. We reinterpret the hierarchy construction as a two-step procedure: stacking with a second FQH state and condensing a condensable algebra of bosons. This two-step procedure can be applied to both abelian and non-abelian FQH states, and it does not require calculations with a wavefunction. We show this construction  reproduces the hierarchies for the Laughlin and Pfaffian states, and can  be applied further to propose hierarchies for various non-abelian FQH states.
\end{abstract}

\maketitle

\setcounter{tocdepth}{1} 
{\small \tableofcontents }

\section{Introduction}

Fractional quantum Hall (FQH) systems can realize many different types of topological orders, 
with excitations carrying fractional charge and fractional statistics. The simplest FQH state occurs at filling fraction $\nu=1/3$, and can be described by the Laughlin wavefunction\cite{L8395}
\begin{equation}
\label{laughlin}
\Psi_{\nu=1/3} = \chi_1^3(\{z_i\}).
\end{equation}

Here, $\chi_n(\{z_i\})$ is the many-fermion wave function of $n$-filled Landau levels, with $\{z_i\}$ labeling the electron positions.
non-abelian topological orders have also been observed in FQH systems. Ising anyons can be realized in the Pfaffian state, described by the wavefunction\cite{moore1991}
\begin{equation}
\label{Pfwv}
\Psi_{\nu=1/2} = \mathcal{A}_\text{anti-symm}\big( 
\frac{1}{z_1-z_2}
\frac{1}{z_3-z_4}
\cdots
\big)\chi_1^2(\{z_i\}).
\end{equation}

Similar non-abelian Ising anyons can be realized in the wavefunction\cite{wen1991,blok1992,W9927,TW230809702}
\begin{equation}
\label{SU22}
\Psi_{\nu=1/2} = \chi_1(\{z_i\}) \chi_2^2(\{z_i\}),
\end{equation}
which has excitations described by the $SU(2)_2$ modular tensor category (MTC). Fibonacci anyons, which are more powerful for topological quantum computation, can be realized by more complicated wavefunctions beyond paired quantum hall
states, such as \cite{wen1991,blok1992,W9927,TW230809702}
\begin{equation}
\label{SU32}
\Psi_{\nu=2/3} = \chi_2^3(\{z_i\}),
\end{equation}
which has non-abelian excitations described by the $SU(2)_3$ MTC. 

The hierarchy construction organizes many of the observed FQH states into
families\cite{haldane1983,halperin1984}. Given a parent FQH state $\eC$ at filling
$\nu$, the hierarchy construction predicts daughter states
$\{\eC_n\}$ at filling fractions $\{\nu_n\}$ that stabilize when the
filling fraction is tuned slightly away from $\nu$. A physical picture behind the construction is the following: tuning the magnetic field away from its original value gives rise to minimal quasihole or minimal quasiparticle excitations. Minimal quasihole (quasiparticle) excitations are simply the anyons carrying the smallest positive (negative) electric charge. When these excitations form at a finite density, they can "condense"\footnote{In this setting, anyons with fractional topological spin are allowed to condense in the presence of a magnetic field. In this work, we use condensation to specifically refer to boson condensation, with the constraints discussed in Sec.~\ref{scondalg}.} into their own FQH state, that is coupled to the original parent state. The hierarchy construction has been used to explain various observed FQH states that appear close to Laughlin states, such as the $\nu=\frac{2}{5}$ state near the $\nu=\frac{1}{3}$ parent state\cite{haldane1983,halperin1984}.

Given an experimental system with a gapped phase at a particular filling fraction, it may be quite difficult to determine the anyon theory that is actually realized. There are an infinite number of anyon theories consistent with a particular filling. Measuring the abelian or non-abelian statistics of the excitations is very difficult. Measuring chiral central
charge\cite{wen1990chiral} of topological order via thermal Hall
conductance\cite{kane1997,banerjee2018} is also not easy. It is much
easier to measure quantized electric Hall conductance, which is proportional to the filling fraction, with very high accuracy. Because FQH theories with the same filling fraction generally have different minimally charged quasihole and quasiparticicle excitations, they have different daughter states with different filling fractions. Thus, measuring filling fractions of the daughter states is an effective and practical way to determine the topological order in a given FQH state. This has been applied to give evidence for whether experimental systems at filling fraction $\frac{1}{2}$ (up to an integer) realize the Pfaffian state or the anti-Pfaffian state\cite{zibrov2017,huang2022,singh2023,hu2024}.  
As mentioned above, the physical process relating a parent state to its daughter states can be viewed as a "condensation" of fractional excitations in the parent state. In other words, the hierarchy construction can be viewed as a generalization of the usual anyon condensation, where only bosonic (or fermionic, in theories with a microscopic fermion) anyons can condense, to condensation of anyons with fractional topological spin in the presence of a magnetic field. In contrast to the usual bosonic anyon condensation, this process produces states with different chiral central charge from their parent state. The procedure for determining the daughter states from a given parent state relies on the use of a Lagrangian or wavefunction (see Ref.~\cite{hansson2017} for a review). It is most well developed for abelian theories \cite{haldane1983,halperin1984} and abelian condensation in non-abelian theories \cite{lan2017,bonderson2008}. non-abelian anyon condensation for the Pfaffian and anti-Pfaffian states was discussed in Ref.~\cite{levin2009}, based on
many-body wave functions. However, a general theory of the hierarchy construction, including for states without known wavefunction descriptions such as the PH-Pfaffian, is still lacking.

\begin{figure}[tb]
   \centering
   \includegraphics[width=.7\columnwidth]{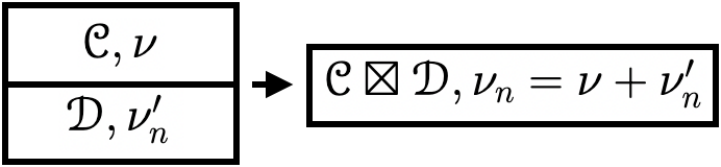} 
   \caption{We construct FQH hierarchies from a parent state $\eC$ at filling fraction $\nu$ by stacking with another theory $\eD$ with filling fraction $\nu_n'$ such that the minimal quasihole/quasiparticle in $\eC$ can bind with an anyon in $\eD$ to form a condensable charge neutral boson that is an element of the condensable algebra $\mathcal{A}$. After condensation of $\mathcal{A}$, the resulting topological orders are the hierarchy states, at filling fraction $\nu+\nu_n'$.}
   \label{fig:fig1}
   \end{figure}
   
In this paper, we use the algebraic theory of $U(1)$ enriched topological order to view the hierarchy construction as
boson condensation. We split the procedure of condensing (non-bosonic) anyons in FQH states into two simpler steps: stacking with a second FQH state $\eD$ and condensing a condensable algebra in $\eC\boxtimes\eD$. This allows us to use well-developed tools in the mathematics literature for condensable algebras\cite{davydov2013,kong2014,etingof2016,cong2017}. Our approach covers both abelian FQH states and non-abelian FQH states in unified and
systematic way, without the need for a wavefunction or Lagrangian. 

\section{Hierarchy construction via boson condensation}

We will first provide a brief summary of our hierarchy construction, and we will work through a simple illustrative example. While our construction is completely general for all parent states $\eC$, we will also give a wavefunction derivation of the construction in Appendix~\ref{swavefunction}, that applies when the FQH state has a known CFT description. This section uses some concepts related to topological order and condensable algebras; these are reviewed in Sec. ~\ref{sreview}.

\subsection{Summary of the hierarchy construction}\label{ssummary}

Given a parent FQH state $\eC$, we obtain daughter states induced by abelian or
non-abelian anyon condensation using a two-step construction:
\begin{enumerate}
\item First, we use the electron excitations in the parent
state $\eC$ to form a topologically ordered state $\eD$ in the presence of the excess magnetic field.
We choose $\eD$ carefully, such that
there exists a condensable algebra $\mathcal{A}$ in $\eC\boxtimes\eD$ satisfying various properties as described later in this section. Furthermore, $\eD$ must have a filling fraction $\nu_n'$ of the correct sign for the quasihole or quasiparticle hierarchy.
\item Second, we condense $\mathcal{A}$ to obtain a new theory $\eC\boxtimes \eD/\mathcal{A}$ with the same chiral central charge as $\eC\boxtimes\eD$ and filling fraction $\nu_n=\nu+\nu_n'$. Note that $\mathcal{A}$ is in general not Lagrangian (maximal), so the resulting daughter states are not trivial.
\end{enumerate}

As will be apparent in the example in Section~\ref{sex1}, the $n$ labeling $\{\eC_{n}\}$ comes from the choice of $\eD$. The condensable algebra $\mathcal{A}$ must satisfy several properties:\footnote{The mathematical framework of condensable algebras is most well-developed for bosonic topological orders. To apply this framework for fermionic systems, we will use minimal modular extension, as explained in Sec.~\ref{scondferm} and Appendix~\ref{sfermionic}.}
\begin{itemize}
\item For quasihole hierarchies, it must include a minimal quasihole, fused with an anyon in $\eD$ such that the composite particle is bosonic. Similarly for quasiparticle hierarchies, it must include a minimal quasiparticle, fused with an anyon in $\eD$ such that the composite particle is bosonic.\footnote{As explained in Sec.~\ref{scondferm}, because we will use the convention where we give the physical fermion a label distinct from the vacuum, we condense only bosons.} In Sec.~\ref{sbondsling}, we relax this constraint, and derive hierarchies based on condensation of non-minimal quasiholes and quasiparticles, recovering the results from Ref.~\cite{bonderson2008} for the Pfaffian state. In some cases, there are multiple distinct minimal quasiholes and quasiparticles (see i.e. Sec.~\ref{ssu23} and Sec.~\ref{sz3}). Different choices of minimal quasiholes and quasiparticles lead to different hierarchy states.
\item The anyons in $\mathcal{A}$ must all carry zero electric charge.\footnote{We will show that even integer charge is also acceptable because even integer charge can be canceled by stacking with an appropriate layers of quantum spin Hall systems. However, in the examples in this paper, we will always assume that the anyons in $\mathcal{A}$ carry zero electric charge for simplicity.} 
\end{itemize}

The first constraint is related to the spin of the anyons we condense, while the second constraint is related to $U(1)$ fractionalization. Additional symmetries like spatial symmetries may lead to additional constraints (see Sec.~\ref{sdiscussion} for some comments on shift). 

Both of the above two steps are mathematically consistent. However in a real physical system, the process of going from a parent state to a daughter state likely looks more like the single step condensation procedure described in the existing literature. In the examples we will work out, $\eD$ tends to be quite complicated, and unlikely to be stable as a standalone FQH phase in a real system of electrons interacting via the Coulomb interaction. Therefore, the two-step procedure may be viewed as a computational tool rather than an accurate description of the microscopic physics. We comment more on the physical interpretation of our hierarchy construction in Sec.~\ref{sdiscussion}.

Note that whenever we write $\eC$ and $\eD$, we mean not only an anyon theory but also its $U(1)$ symmetry fractionalization pattern. As a $U(1)$ enriched topological order, $\eD$ also has a filling fraction $\nu_n'$. Because we only condense charge zero bosons, we obtain daughter states with filling, as mentioned earlier, given by simply $\nu_{n}=\nu+\nu_n'$. We use a $\eD$ with opposite chirality in the $U(1)$ factor as $\eC$ to construct quasihole hierarchies and a $\eD$ with the same chirality in the $U(1)$ factor as $\eC$ to construct quasiparticle hierarchies.

It may seem that the above construction is too general. There is an infinite number of $U(1)$ enriched topological orders $\eD$ that we can stack with $\eC$. This produces a vast array of mathematically consistent daughter states.  The requirement that $\eC\boxtimes \eD$ contains a condensable charge 0 boson $\v a\v b$ where $\v a$ is the minimal quasihole or quasiparticle of $\eC$ and $b\in\eD$ narrows down the options significantly. In fact, we would like the entire condensable algebra to be generated by $\v a\v b$ (in the sense that every anyon in $\mathcal{A}$ can be found in the fusion product of $\v a\v b$ with itself a finite number of times), because $\v a$ is the lowest energy excitations of $\eC$ when the magnetic field is slightly altered. We have four more physically motivated guiding principles. First, we search for condensations that give daughter states with the smallest rank, because these are more stable and likely to be realized by electrons interacting by the Coulomb interaction. This indicates that the chiral central charge of $\eC\boxtimes\eD$, which is invariant under the condensation, should be a rather simple fraction. Second, we look for filling fractions $\nu_n'$ that have relatively small absolute value, to be consistent with the picture that these daughter states can be obtained by shifting the filling slightly away from $\nu$. Third, we look for $\eD$ that has small chiral central charge, so that the change in chiral central charge between a parent state and its daughter states is small. Finally, we also prefer $\eD$ that is fully chiral or fully antichiral, because in these cases we can often use the CFT ansatz to obtain holomorphic/antiholomorphic wavefunctions\cite{moore1991,hansson2017}.

We note that condensable algebras are rather complex objects; one of the pieces of data defining a condensable algebra (the algebra morphism) is a solution to a nonlinear consistency equation similar to the pentagon equation (see Sec.~\ref{sreview}). In the following, we will obtain daughter states by proposing condensable algebras. We will check that these algebras satisfy some
conditions that strongly indicate that they may be valid condensable algebras,
but we do not solve the nonlinear consistency equation to guarantee that they are valid condensable algebras.

\subsection{Example: hierarchy
construction for $\nu=\frac{1}{3}$ FQH phase}\label{sex1}
Let us apply the above approach to obtain the daughter/ states of the simplest FQH phase, the
$\nu=\frac{1}{3}$ abelian FQH phase. A Lagrangian density for this phase is given by $U(1)_3$ Chern-Simons theory:
\begin{equation}
\mathcal{L}=-\frac{3}{4\pi}a_{\mu}\partial_{\nu}a_{\lambda}\epsilon^{\mu\nu\lambda},
\end{equation}
where $a$ is an emergent dynamical $U(1)$ gauge field. The theory is described by abelian anyons with a $\mathbb{Z}_6$ fusion rule. We label them by $\v a^k$ for $k\in[0,5]$:
\begin{equation}
\{\v 0,\v a,\v a^2,\v a^3,\v a^4,\v a^5\}.
\end{equation}

Here, $\v a^3$ is the transparent fermion; it braids trivially with all other excitations. We use the notation $\v 0$ for vacuum to avoid confusion with $\v 1$, which refers to the spin 1 representation of $SU(2)$, when we later discuss $SU(2)_m$. The topological spins and charges of the anyons are given by
\begin{equation}\label{spincharge}
\theta_{\v a^k}=e^{2\pi i k^2/6}\qquad q_{\v a^k}=\frac{k}{3}.
\end{equation}

It is easy to check using (\ref{spincharge}) that indeed $\v a^3$ has trivial braiding statistics with all other excitations in the theory.

To obtain the quasiparticle hierarchy states, we would like a theory $\eD$ that has an anyon that can form a boson together with $\v a^{-1}$. The simplest theory with such an anyon is $U(1)_m$.\footnote{More generally, cyclic anyon theories can be labeled by coprime integers $m,p$, where $p=1$ for $U(1)_m$. These are also known as "minimal abelian TQFTs"\cite{hsin2019}, and can be described by multi-component $U(1)$ Chern-Simons theory. The spins of a theory labeled by $m,p$ is given by $e^{2\pi i p l^2/(2m)}$ where $l\in[0,m-1]$ for $pm$ even (these are bosonic theories) and $l\in[0,2m-1]$ otherwise. In Sec.~\ref{slaughlin}, we discuss stacking with these more general abelian theories. We find that they are less likely to result in observed hierarchy states because the shift in the filling fraction $\nu_n'$ is much larger, and the chiral central charges of the hierarchy states are also larger.} Note that we use $U(1)_m$ for the quasiparticle hierarchy because it has positive filling fraction. As a result, the filling fractions of the daughter states $\nu_n$ will be greater than $\nu$, as expected for the quasiparticle hierarchy. The anyons $\v b^l$ of $U(1)_m$ have topological spin $\theta_{\v b^l}=e^{2\pi il^2/(2m)}$, so we need to solve the equation
\begin{equation}\label{lmn}
\frac{l^2}{2m}=-\frac{1}{6}+n=\frac{6n-1}{6},
\end{equation}
where $n$ is an integer. This ensures that $\v a^{-1}\v b^l$ is a boson. In addition, we need $\v a^{-1}\v b^{l}$ to have zero charge. For a $U(1)$ symmetry fractionalization pattern in $\eD$ where the fractional charges are assigned by braiding with $\v b^v\equiv \v v$ (see Sec.~\ref{sreview}; $\v v$ from this point onward will always be used to refer to the vison, which assigns anyons their fractional $U(1)$ charge), this gives the constraint
\begin{equation}
\frac{lv}{m}=\frac{1}{3}.
\end{equation}

Putting this together with (\ref{lmn}) gives
\begin{equation}
m=3(6n-1)v^2\qquad l=(6n-1)v
\end{equation}

We see that for $m$ odd (so $\eD$ is fermionic), $v$ must also be odd. This is consistent with the requirement that the transparent fermion must carry odd integer charge. The filling fraction can be computed from the relation $\theta_{\v v}=\theta_{\v b^v}=e^{\pi i \nu}$. This gives
\begin{equation}
\nu_n'=\frac{1}{3(6n-1)}.
\end{equation}

For example, for $n=1$, we have $\nu_1'=\frac{1}{15}$. The resulting daughter state has filling fraction
\begin{equation}
\nu_{\mathrm{tot}}=\frac{1}{3}+\frac{1}{15}=\frac{2}{5},
\end{equation}
as expected of the first quasiparticle daughter state of $U(1)_3$. The simplest example of $\eD$ is just the $U(1)_{15}$ anyon theory with fractional charge assigned by braiding with the generator; this clearly has filling fraction $\frac{1}{15}$. $l=5v$ and $m=15v^2$, so we have the condensable algebra 
\begin{equation}
\mathcal{A}=\v 0+\v a^{-1}\v b^{5v}+\dots+\v a^{-3 v+1}\v b^{15 v^2-5v}.
\end{equation}

Note that we can condense $\v a^{-3 v}\v b^{15 v^2}$ because it is a boson that braids trivially with everything else. This condensable algebra has quantum dimension $d_{\mathcal{A}}=3v$, while $\eC\boxtimes\eD$ has total quantum dimension $\sqrt{2\times 45v^2}$, so the topological phase after condensation has total quantum dimension $\frac{D}{d_{\mathcal{A}}}=\sqrt{10}$.

Indeed, after condensing $\mathcal{A}$ the anyons that remain form an abelian topological order generated by an anyon of order $10$. The anyons that remain are those that braid trivially with all the condensed anyons. Up to equivalence under fusion with the condensed anyons, we can choose these to be generated by $\v b^{-3v}$. This anyon has topological spin $e^{2\pi i 3/10}$ and charge $\frac{-1}{5}$. These topological spins and fractional charges match those of the hierarchy state derived from the Lagrangian perspective\cite{haldane1983}, with 
\begin{equation}\label{kmatrixcond}
\mathcal{L}=-\frac{1}{4\pi}K_{IJ}a_{I\mu}\partial_{\nu}a_{j\lambda}\epsilon^{\mu\nu\lambda}+\frac{1}{2\pi}q_IA_{\mu}\partial_{\nu}a_{I\lambda}e^{\mu\nu\lambda},
\end{equation}
with integer $K$ matrix and charge vector
\begin{equation}
K=\begin{pmatrix}3 & -1\\ -1 & 2\end{pmatrix}\qquad \v q=\begin{pmatrix} 1 \\ 0\end{pmatrix},
\end{equation}
and minimal charge quasiparticle labeled by integer vector $\v l^T=\begin{pmatrix} 0 & 1\end{pmatrix}$. In Sec.~\ref{slaughlin}, we explicitly relate the $\mathcal{A}$ condensed theory to (\ref{kmatrixcond}) by finding a $GL(N,\mathbb{Z})$ change of basis that maps between the two theories (each stacked with trivial quantum spin Hall states).

More generally, we find daughter states with filling
\begin{equation}
\nu_{n}=\frac{1}{3}+\frac{1}{3(6n-1)}=\frac{2n}{6n-1},
\end{equation}
which matches the results of Refs.~\cite{haldane1983,halperin1984} with the choice $p_1=3$ and $p_2=2n$.

\section{MTCs, sMTCs, and condensable algebras}
In this section, we review some relevant aspects of $U(1)$ symmetry enriched topological orders and condensable algebras. Much of the work on condensable algebras assume that the topological order is bosonic, meaning the braiding is nondegenerate. Because we are concerned with FQH states, which are fermionic, we need to adapt the work on bosonic systems to fermionic systems. We give two approaches for studying condensable algebras in fermionic systems in Sec.~\ref{smme} and Sec.~\ref{sfermioniccond}. Note that bosonic topological orders are described by unitary modular tensor categories (MTCs), while fermionic topological orders are described by supermodular tensor categories (sMTCs).

\subsection{Review of bosonic and fermionic anyon topological order}\label{sreview}
A topological order $\eC$ consists of several pieces of data. The first piece of data is a finite list of simple objects (anyons) $\{\v a\}$. The second piece of data is the set of fusion rules for $\{\v a\}$, given by integer fusion coefficients $\{N_{\v a\v b}^{\v c}\}$:
\begin{equation}
\v a\times\v b=\sum_{\v c}N_{\v a\v b}^{\v c}\v c
\end{equation}

Each anyon $\v a$ has a conjugate anyon $\bar{\v a}$ with $N_{\v a\bar{\v a}}^{\v 0}=1$, where $\v 0$ is the vacuum. The fusion rules determine a quantum dimension $d_{\v a}$ for each anyon, with
\begin{equation}
d_{\v a} d_{\v b}=\sum_{\v c}N_{\v a\v b}^{\v c}d_{\v c},
\end{equation}
and $d_{\v 0}=1$. The total quantum dimension is $\mathcal{D}=\sqrt{\sum_{\v a\in\eC}d_{\v a}^2}$. Fusion rules and quantum dimensions will be important for studying condensable algebras. For simplicity of notation, we will assume that $N_{\v a\v b}^{\v c}\in\{0,1\}$ for all $\v a,\v b,$ and $\v c$. This will be sufficient for the examples in this paper; the extension to more general fusion coefficients is straightforward. 

Another piece of data describing a topological order is the $F$ and $R$ symbols, which describe the associativity and braiding of the anyons respectively. We will not need the $F$ symbol in this work, but we will use the $R$ symbol:
\begin{equation}
\includegraphics{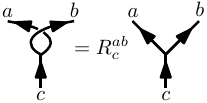}.
\end{equation}

The $R$ symbol, together with the quantum dimensions, determines the topological spins of the anyons:
\begin{equation}\label{rspin}
\theta_{\v a}=\theta_{\bar{\v a}}=\sum_{c}\frac{d_{\v c}}{d_{\v a}}R^{\v a\v a}_{\v c}.
\end{equation}

It follows from (\ref{rspin}) that
\begin{equation}
R^{\v a\v b}_{\v c}R^{\v b\v a}_{\v c}=\frac{\theta_{\v c}}{\theta_{\v a}\theta_{\v b}}.
\end{equation}

The braiding properties of the anyons can be packaged into two important gauge invariant quantities: the $S$ and $T$ matrices:
\begin{equation}
S_{\v a\v b}=\frac{1}{\mathcal{D}}\sum_{\v c}N_{\bar{\v a}\v b}^{\v c}\frac{\theta_{\v c}}{\theta_{\v a}\theta_{\v b}}d_{\v c},
\end{equation}
\begin{equation}
T_{\v a\v b}=\theta_{\v a}\delta_{\v a\v b}.
\end{equation}
In bosonic theories, these two matrices are nondegenerate.

The final piece of data defining a topological order is the chiral central charge. In bosonic theories, the chiral central charge is determined modulo 8 by the algebraic properties of the anyons:
\begin{equation}
    \frac{1}{\mathcal{D}}\sum_{\v a\in\mathcal{C}}d_{\v a}^2\theta_{\v a}=e^{\frac{2\pi ic_-}{8}}.
\end{equation}

However, the physical chiral central charge is defined on the real numbers, not modulo 8, and is proportional to the thermal Hall conductance.

We will mostly work with fermionic topological orders. In these theories, the $S$ matrix is degenerate because there is a fermion that braids trivially with all other anyons in the theory. Furthermore, the chiral central charge is only defined modulo 1/2 by the algebraic data, because the invertible p-wave superconductor has chiral central charge 1/2 and no anyons.

\subsection{$U(1)$ symmetry fractionalization}
In the presence of $U(1)$ symmetry, the anyons are also labeled by a fractional charge according to the $U(1)$ symmetry fractionalization pattern. Specifically, the $U(1)$ symmetry action localized on an anyon $\v a$ may have a projective representation
\begin{equation}
U_{\v a}(\pi)U_{\v a}(\pi)=e^{2\pi i q_{\v a}}U_{\v a}(0),
\end{equation}
where $q_{\v a}$ is the fractional charge of $\v a$. In order for the charge assignments to be consistent with the fusion rules of the anyons, it must be possible to write $e^{2\pi i q_{\v a}}$ as the braiding phase of $\v a$ with a particular abelian anyon $\v v$. Therefore, the fractional charge $q_{\v a}$ of each anyon $\v a\in\eC$ is given by the braiding phase of $\v a$ with $\v v$. Different choices of $\v v$ correspond to different fractionalization patterns.

In fermionic theories, we must ensure that the transparent fermion has odd integer charge. This puts constraints on the choice of $\v v$. For example, in the $U(1)_3$ theory (\ref{spincharge}), we cannot choose $\v v=\v a^2$ because that would assign the transparent fermion, $\v a^3$, even integer charge.

The filling fraction can be read off from the topological spin of $\v v$:
\begin{equation}
    \theta_{\v v}=e^{\pi i\nu}.
\end{equation}

This comes from the fact that the filling fraction gives the charge accumulated at a $2\pi$ flux insertion. Here, a $2\pi$ flux insertion hosts the abelian anyon $\v v$, so the charge is simply given by the charge of $\v v$, which is determined by the braiding phase of $\v v$ with itself.

\subsection{Condensable algebras}\label{scondalg}
A condensable algebra\cite{davydov2013,kong2014,etingof2016,cong2017} in a (bosonic) topological order is a composite anyon of the form
\begin{equation}
\mathcal{A}=\oplus_{\v a\in\eC}n_{\v a} \v a
\end{equation}
where $n_{\v a}$ is a nonnegative integer describing the number of inequivalent ways to annihilate $\v a$ at the boundary. It therefore gives the dimension of a boundary condensation space $V^{\v a}$. $\mathcal{A}$ must include $\v 0$, with $n_{\v 0}=1$. 

$\mathcal{A}$ describes a collection of bosons that can be simultaneously condensed in the bulk and boundary of the topological order. The quantum dimension of $\mathcal{A}$ is given by 
\begin{equation}
   d_{\mathcal{A}}=\sum_{\v a\in\eC}n_{\v a}d_{\v a} .
\end{equation}

When $d_{\mathcal{A}}=\mathcal{D}$, the condensable algebra is Lagrangian and specifies a boundary between $\eC$ and the vacuum. When $d_{\mathcal{A}}<\mathcal{D}$, the condensable algebra instead describes a boundary between $\eC$ and another topological order, which is the topological order that results from condensing $\mathcal{A}$ in $\eC$. 

A condensable algebra must also come with an $M$ symbol, which compares annihilating $\v a\times \v b$ (with boundary condensation space $V^{\v a}\otimes V^{\v b}$) at the boundary and their fusion products at the boundary (with boundary condensation space $V^{\v c}$). The $M$ symbol is not unitary:\cite{cong2017}
\begin{equation}\label{ninequ}
n_{\v a}n_{\v b}\leq\sum_{\v c}N_{\v a\v b}^{\v c}n_{\v c}.
\end{equation}
The inequality in (\ref{ninequ}) is generally not saturated for condensable algebras of non-abelian topological orders. 

The $M$ symbol must satisfy a nonlinear pentagon-like equation with the $F$ symbol, as well as a consistency equation with the $R$ symbol that leads to the following condition:
\begin{equation}\label{rconstraint}
R^{\v a\v b}_{\v c}R^{\v b\v a}_{\v c}=1,
\end{equation}
if $\v a,\v b,$ and $\v c$ are all in $\mathcal{A}$. Note that for $\v a,\v b\in\mathcal{A}$, their braiding does not need to be trivial in every fusion channel. However, there must be trivial braiding in at least one fusion channel $\v c$, so that $n_{\v c}\neq 1$. Otherwise, (\ref{ninequ}) requires that $n_{\v a}=n_{\v b}=0$.

In this work, we will propose many condensable algebras. We will not show that they have valid $M$ symbols because this is a rather difficult task. In Appendix~\ref{sstmat}, we discuss an easier necessary but not sufficient condition for a condensable algebra $\mathcal{A}$ to describe a valid gapped domain wall between $\eC$ and $\eC/\mathcal{A}$. For the condensable algebras proposed in this paper, we will actually not have to check the condition in Appendix~\ref{sstmat} either, because they all compose of sequences of diagonal condensations (of the form $\sum a\bar{a}$, where $\bar{a}$ is the time reversal of $a$ and the sum is taken over all of the simple objects in an MTC) and abelian condensations. However, more complex condensations may need the additional checks presented in Appendix~\ref{sstmat}. One very useful property derived from the conditions in Appendix~\ref{sstmat} is the following condition on the total quantum dimension of a theory before and after condensing $\mathcal{A}$:
\begin{equation}\label{dimchange}
\frac{\mathcal{D}_{\eC}}{d_{\mathcal{A}}}=\mathcal{D}_{\eC/\mathcal{A}}.
\end{equation} 

The condensable algebras are also constrained by the $U(1)$ symmetry fractionalization. In particular, the condensed anyons must all carry trivial charge, meaning they braid trivially with $\v v$:
\begin{equation}
e^{2\pi i q_{\v a}}=\theta_{\v a,\v v}=1\qquad\forall \v a\in\mathcal{A}.
\end{equation}

\subsubsection{Condensable algebras for fermionic theories}\label{scondferm}
Strictly speaking, the above formalism is developed for bosonic topological orders only, not fermionic ones. In abelian systems, condensations are labeled instead by condensable subgroups. This is just a subgroup of anyons that braid trivially with each other. The extension of condensable subgroups to fermionic theories is straightforward\cite{levin2013}. Note that in Ref.~\cite{levin2013}, the physical fermion was identified with the vacuum, so condensable subgroups included anyons with fermionic statistics. These anyons pair with the physical fermion to condense as bosons. In this work, we will give the fermion a distinct anyon label from the vacuum, so our condensable subgroups will always contain only bosons.

There are two approaches for more rigorously defining condensable algebras for fermionic theories. We explain these two approaches in Appendix~\ref{sfermionic}. The approach we take in the rest of this paper is to first obtain a minimal modular extension of $\eC$ to $\eC^{(b)}$, and then perform the condensation in the bosonic theory. A minimal modular extension is a theory that contains all the anyons in the original theory, together with additional anyons that are fermion parity fluxes. The fluxes braid nontrivially with the formerly transparent fermion, so they make the theory modular. In other words, minimal modular extensions are the minimal extensions of a fermionic theory to a bosonic, nondegenerate theory. For every fermionic topological order, there are 16 minimal modular extensions. For example, the minimal modular extensions of the trivial fermionic theory $\{1,f\}$, where $f$ is a fermion, are simply Kitaev's 16-fold way\cite{kitaev2006}. The total quantum dimension of a minimal modular extension $\eC^{(b)}$ is related to that of the original fermionic theory $\eC$ by
\begin{equation}
    D_{\eC^{(b)}}=\sqrt{2}D_{\eC},
\end{equation}
if we count the transparent fermion as an anyon in $\eC$.

In summary, we first perform a modular extension of $\eC$ to $\eC^{(b)}$, and then we stack with $\eD$, and finally we condense a condensable algebra in $\eC^{(b)}\boxtimes \eD$. Finally, to return to a fermionic theory, we condense the transparent fermion, projecting out anyons that braid nontrivially with it. This removes the fluxes, undoing the modular extension. We must ensure that the hierarchy states are independent of the modular extension. In this work, all of the condensable algebras contain only anyons from the fermionic theory $\eC$; they do not contain any fermion parity fluxes. It follows that the hierarchy states do not depend on the choice of extension. The extension serves merely as a formality to make the condensation more rigorous.  

\subsubsection{Constraints from condensable algebra}\label{sconstraints}
As discussed in Sec.~\ref{ssummary}, there is a vast array of different theories $\eD$ that we can try to stack onto any given $\eC$, so it seems that there are too many different daughter states for a given parent state. However, even if $\eC\boxtimes\eD$ contains an anyon $\v a\v b$ that is a boson, it may not have a nontrivial condensable algebra that satisfies conditions like (\ref{ninequ}), (\ref{scaint}), and (\ref{tconstraint}).

For example, suppose that we tried to obtain the daughter states for the Pfaffian state using $\eD$ that is a abelian. A minimal modular extension of Pfaffian is simply $\mathrm{Ising}\boxtimes U(1)_8$. The minimal charge quasiparticle is $\sigma \v a$ where $\v a$ generates the $U(1)_8$. Suppose that we can pair $\sigma \v a$ with an anyon $\v b\in\eD$ such that $\sigma \v a\v b$ is a boson. Then 
\begin{equation}
\theta_{\v a\v b}=\theta_{\sigma}^{-1}=e^{-2\pi i/16}.
\end{equation}

Fusing two of these bosons gives
\begin{equation}\label{sigmafusion}
\sigma\v a\v b\times\sigma\v a\v b=(1+\psi)\v a^2\v b^{2}.
\end{equation}
The topological spin of $\v a^2\v b^{2l}$ is 
\begin{equation}
\theta_{\v a^2\v b^{2}}=\left(\theta_{\v a\v b}\right)^4=-i.
\end{equation}

It follows that none of the anyons in the fusion product (\ref{sigmafusion}) are bosons. Therefore, none of the anyons in the fusion product can be in $\mathcal{A}$. According to (\ref{ninequ}), $\sigma \v a\v b$ also cannot be in $\mathcal{A}$. 

Therefore, we cannot obtain a condensable algebra with the minimal charge quasiparticle of the (modular extension of the) Pfaffian state by simply stacking the theory with any abelian theory. In Sec.~\ref{spfaff}, we obtain condensable algebras by stacking the Pfaffian state with non-abelian topological orders.

\begin{table*}[t]
\caption{
Some examples of FQH states $\eC$ with filling fraction $\nu$ and their daughter states $(\eC\boxtimes \eD)/\cA$ with filling fraction $\nu_n$, obtained by condensing minimally charged excitations. Some parent states have two different minimally charged excitations, which lead to two different hierarchies. $c_-$ is the chiral central charge of the hierarchy states. The bottom set of rows correspond to the $k=3$ Read Rezayi states at filling $\nu=3/5$. In last three groups, the first two daughter states may be more stable, since the condensing anyons give rise to chiral edge states with smaller central charges.}
\label{table1}
\begin{tabular}
{ |p{6.3cm}||p{5cm}|p{0.5cm}|p{1.6cm}|p{1.7cm}| p{1cm}|  }
 \hline
 $\eC$ & $\eD$  & $\nu$ & $\nu_{n}$ & $\nu_{ n=1}$  & $c_-$ \\
 \hline
 \hline
 $U(1)_m$ ($\chi_1^m$, quasihole)   & $U(1)_{-m(2mn+1)}$  & $\frac{1}{m}$ & $\frac{2n}{2mn+1}$ &   $\frac{2}{7}$ $(m=3)$ & $0$\\
 $U(1)_m$ ($\chi_1^m$, quasiparticle) & $U(1)_{m(2mn-1)}$ & $\frac{1}{m}$ &  $\frac{2n}{2mn-1}$  & $\frac{2}{5}$  $(m=3)$ & $2$\\
 \hline
 Pfaffian (quasihole) & $\overline{\mathrm{Ising}}\boxtimes U(1)_{-8(16n+1)}/\mathbb{Z}_2$ &$\frac{1}{2}$ & $\frac{8n}{16n+1}$ & $\frac{8}{17}$ &  $0$\\
  Pfaffian (quasiparticle) & $\mathrm{Ising}\boxtimes U(1)_{8(16n-3)}/\mathbb{Z}_2$ & $\frac{1}{2}$ & $\frac{8n-1}{16n-3}$ & $\frac{7}{13}$ &  $3$\\

\hline
 Bosonic Pfaffian (quasihole) & $\overline{\mathrm{Ising}}\boxtimes U(1)_{-4(8n+1)}$    & $1$ & $\frac{8n}{8n+1}$ & $\frac{8}{9}$ & $0$ \\
  Bosonic Pfaffian (quasiparticle) & $\mathrm{Ising}\boxtimes U(1)_{2(4n-1)}$ & $1$ & $\frac{8n-1}{8n-2}$ & $\frac{7}{6}$ & $3$\\
 \hline
 PH-Pfaffian (quasihole) & $\mathrm{Ising}\boxtimes U(1)_{-8(16n+1)}/\mathbb{Z}_2$    & $\frac{1}{2}$ & $\frac{8n}{16n+1}$ & $\frac{8}{17}$ & $0$ \\
  PH-Pfaffian (quasiparticle) & $\overline{\mathrm{Ising}}\boxtimes U(1)_{8(16n+1)}/\mathbb{Z}_2$ & $\frac{1}{2}$ & $\frac{8n+1}{16n+1}$ & $\frac{9}{17}$ & $1$\\
  PH-Pfaffian (quasihole)  &   $\overline{\mathrm{Ising}}\boxtimes U(1)_{-8(16n-1)}/\mathbb{Z}_2$     &  $\frac{1}{2}$  &  $\frac{8n-1}{16n-1}$  &  $\frac{7}{15}$  &  $-1$ \\
  PH-Pfaffian (quasiparticle) &  $\mathrm{Ising}\boxtimes U(1)_{8(16n-1)}/\mathbb{Z}_2$  &  $\frac{1}{2}$ & $\frac{8n}{16n-1}$ & $\frac{8}{15}$  & $2$\\
 \hline
 $SU(2)_2\boxtimes U(1)_8/\mathbb{Z}_2$ ($\chi_1\chi_2^2$, quasihole) & $\overline{\mathrm{Ising}}\boxtimes U(1)_{-8(16n+3)}/\mathbb{Z}_2$ & $\frac{1}{2}$ & $\frac{8n+1}{16n+3}$ & $\frac{9}{19}$ &  $1$\\
 $SU(2)_2\boxtimes U(1)_8/\mathbb{Z}_2$ ($\chi_1\chi_2^2$, quasiparticle) & $\mathrm{Ising}\boxtimes U(1)_{8(16n-5)}/\mathbb{Z}_2$ & $\frac{1}{2}$ & $\frac{8n-2}{16n-5}$ & $\frac{6}{11}$ &  $4$\\[2mm]
   $SU(2)_2\boxtimes U(1)_8/\mathbb{Z}_2$ ($\chi_1\chi_2^2$, quasihole) & $SU(2)_{-2}\boxtimes U(1)_{-8(16n+1)}/\mathbb{Z}_2$ & $\frac{1}{2}$ & $\frac{8n}{16n+1}$ & $\frac{8}{17}$ &  $0$\\
 $SU(2)_2\boxtimes U(1)_8/\mathbb{Z}_2$ ($\chi_1\chi_2^2$, quasiparticle) & $SU(2)_2\boxtimes U(1)_{8(16n-7)}/\mathbb{Z}_2$ & $\frac{1}{2}$ & $\frac{8n-3}{16n-7}$ & $\frac{5}{9}$ &  $5$\\
 \hline
  $SU(2)_3\boxtimes U(1)_6/\mathbb{Z}_2$ ($\chi_2^3$, quasihole) & $U(1)_{-3(6n+5)}$ & $\frac{2}{3}$ & $\frac{4n+3}{6n+5}$ & $\frac{3}{5}$  &  $\frac{9}{5}$\\
 $SU(2)_3\boxtimes U(1)_6/\mathbb{Z}_2$ ($\chi_2^3$, quasiparticle) & $U(1)_{3(6n-5)}$  & $\frac{2}{3}$  & $\frac{4n-3}{6n-5}$ & $1$ &  $\frac{19}{5}$\\[2mm]
  $SU(2)_3\boxtimes U(1)_6/\mathbb{Z}_2$ ($\chi_2^3$, quasihole) & $\mathrm{Pf}_{-3}\boxtimes U(1)_{-3(6n-1)}$ & $\frac{2}{3}$ & $\frac{12n-3}{18n-3}$ & $\frac{9}{15}$  & $1$ \\
 $SU(2)_3\boxtimes U(1)_6/\mathbb{Z}_2$ ($\chi_2^3$, quasihole) & $SU(2)_{-3}\boxtimes U(1)_{-6(12n+1)}/\mathbb{Z}_2$ & $\frac{2}{3}$ & $\frac{8n}{12n+1}$ & $\frac{8}{13}$  &  $0$\\
 $SU(2)_3\boxtimes U(1)_6/\mathbb{Z}_2$ ($\chi_2^3$, quasiparticle) & $SU(3)_{2}\boxtimes U(1)_{3(2n-1)}$  & $\frac{2}{3}$  & $\frac{4n-3}{6n-5}$ & $1$ &  $7$\\
 \hline
    $\mathbb{Z}_3$ parafermion (quasihole) & $U(1)_{-5(10n+7)}$ & $\frac{3}{5}$ & $\frac{6n+4}{10n+7}$ & $\frac{10}{17}$ &  $\frac{4}{5}$\\
   $\mathbb{Z}_3$ parafermion (quasiparticle)  & $U(1)_{5(10n-7)}$ & $\frac{3}{5}$ & $\frac{6n-4}{10n-7}$ & $\frac{2}{3}$ &  $\frac{14}{5}$\\[2mm] 
  $\mathbb{Z}_3$ parafermion (quasihole) & $\mathrm{Pf}_{-3}\boxtimes U(1)_{-15(30n+1)}$ & $\frac{3}{5}$ & $\frac{18n}{30n+1}$ & $\frac{18}{31}$ &  $0$\\
     $\mathbb{Z}_3$ parafermion (quasiparticle)  & $SU(3)_2\boxtimes U(1)_{5(10n-7)}$ & $\frac{3}{5}$ & $\frac{6n-4}{10n-7}$ & $\frac{2}{3}$ &  $6$\\
 \hline 
\end{tabular}\label{table1}
\end{table*}

\section{Examples}
We now apply our approach to derive daughter states for various FQH states. Our results are summarized in Table ~\ref{table1}.

\subsection{Laughlin states}\label{slaughlin}
We first generalize our example from Sec.~\ref{sex1} to other Laughlin states. Because $\eC$ is abelian, we will choose to stack with $\eD$ that is also abelian,\footnote{One can also consider stacking with $\eD$ that is non-abelian. As discussed in Sec.~\ref{sconstraints}, this can only work if $\eD$ has a non-abelian anyon $\v b$ satisfying $R^{\v b\v b}_{\v c}R^{\v b\v b}_{\v c}=1$ for some $\v c$ in the fusion product $\v b\times\v b$. This is satisfied for some non-abelian sMTCs; we leave a full exploration of possible stackings to future work (see Sec.~\ref{sdiscussion})} we can work with condensable subgroups rather than condensable algebras. Condensable subgroups are well understood in fermionic systems; we will not need to use minimal modular extension in this case. The only difference from the bosonic case is that we allow fermions in the condensable subgroup. However, since we would like to condense charge zero (modulo even integers) objects, we will not include any fermions in our condensable subgroups.

Laughlin states at filling fraction $\nu=\frac{1}{p_1}$ are described by theories $\eC$ of abelian cyclic anyons $\v a^k$ where $k\in[0,2N-1]$. These have topological spin and charge
\begin{equation}
\theta_{\v a^{k}}=e^{\frac{2\pi i k^2}{2p_1}}\qquad q_{\v a^k}=\frac{k}{p_1}.
\end{equation}

The minimal quasiparticle is $\v a^{2p_1-1}=\v a^{-1}$, which has charge $-\frac{1}{p_1}$. For the quasiparticle hierarchy, we stack with another cyclic abelian theory with anyons $\v b^l$ with topological spin and charge
\begin{equation}
\theta_{\v b^{l}}=e^{\frac{2\pi i pl^2}{2m}}\qquad q_{\v b^l}=\frac{lp v}{m}.
\end{equation}

As mentioned in Sec.~\ref{sex1}, $p$ is an integer that is coprime to $m$, and the theory is fermionic if $pm$ is odd. $\eD$ has minimal chiral central charge if $p=1$ and $m$ is odd, so we will proceed with this assumption. In other words, we assume $\eD$ to be described by $U(1)_m$ Chern-Simons theory. $v$ labels the anyon in $\eD$ that determines the fractional charges of the anyons. In order for $\v a^{-1}\v b^l$ to be a boson, we must have
\begin{equation}\label{bosoncond}
\frac{l^2}{2m}=-\frac{1}{2p_1}+n=\frac{2p_1n-1}{2p_1},
\end{equation}
where $n$ is an integer. In order for $\v a^{-1}\v b^l$ to have charge zero, we must have
\begin{equation}\label{chargecond}
\frac{lv}{m}=\frac{1}{p_1}\to l=\frac{m}{vp_1}.
\end{equation}

Putting together (\ref{bosoncond}) and (\ref{chargecond}) with $p=1$ gives
\begin{equation}
\frac{(2p_1n-1)m}{p_1}=\left(\frac{m}{p_1v}\right)^2\to m=p_1(2p_1n-1)v^2.
\end{equation}

Then we can read off from (\ref{chargecond}) that 
\begin{equation}
l=(2p_1n-1)v.
\end{equation}

This theory has filling fraction $\frac{v^2}{m}=\frac{1}{p_1(2p_1n-1)}$. The total filling fraction of the daughter state labeled by $n$ is
\begin{equation}
\nu_{n}=\frac{1}{p_1}+\frac{1}{p_1(2p_1n-1)}=\frac{2n}{2p_1n-1}.
\end{equation}

Relabeling $2n=p_2$ (so $p_2$ is an even integer), we obtain
\begin{equation}
\nu_{p_2}=\frac{p_2}{p_1p_2-1},
\end{equation}
which matches with Refs.~\cite{haldane1983,halperin1984}.\footnote{Note that we can also choose $\v a^{-1}\v b^l$ to have even integer charge:
\begin{equation}
    l=\frac{m}{v}\left(\frac{1}{p_1}+2n'\right)
\end{equation}
where $n'$ is an integer. Solving for $m$ gives
\begin{equation}
    m=\frac{p_1(2p_1n-1)}{(2n'p_1+1)^2}v^2
\end{equation}
The even integer charge can be compensated by quantum spin Hall states. For example, for $n=n'=1$ we get $m=15$ and $v=7$. To compensate for the even integer charge, we can choose $l=\begin{pmatrix} 1 & -5 & -1 & -1\end{pmatrix}$ in (\ref{kmatq}). The filling fraction for $n'>1$ is clearly larger than that for $n=1$ (in fact, the filling fraction is greater than 1 for $n=n'$), so this hierarchy is less likely to be observed. }

The anyons generated by $\v a^{-1}\v b^l$ form a condensable subgroup $\mathcal{A}$. Notice that $\v a^{-1}\v b^l$ is an anyon of order $p_1v$ (once we condense the boson formed by the product of the two transparent fermions), so they form a condensable subgroup of $p_1v$ anyons. The total quantum dimension after condensation is
\begin{equation}
\mathcal{D}_{\eC\boxtimes\eD/\mathcal{A}}=\frac{\sqrt{2p_1^2(p_1p_2-1)v^2}}{p_1v}=\sqrt{2(p_1p_2-1)}.
\end{equation}

The theory is generated by $\v b^{-p_1v}$, which has topological spin $e^{\frac{2\pi ip_1}{2(p_1p_2-1)}}$ and charge $-\frac{1}{p_1p_2-1}$.

For the quasihole hierarchy, we stack with $U(1)_{-m}$ Chern-Simons theory,\footnote{For the same reason as for the quasiparticle hierarchy, we will only consider $p=1$ for elementary abelian anyons labeled by coprime $p,-m$.} which has anyons $\bar{\v b}^l$ with topological spin and charge
\begin{equation}
\theta_{\bar{\v b}^{l}}=e^{-\frac{2\pi i l^2}{m}}\qquad q_{\bar{\v b}^l}=-\frac{l}{m}.
\end{equation}

For $\v a\bar{\v b}^l$ to be a boson, we must have
\begin{equation}
\frac{l^2}{2m}=\frac{2p_1n+1}{2p_1},
\end{equation}
where $n$ is an integer. On the other hand for $\v a\bar{\v b}^l$ to have charge zero, we require
\begin{equation}
l=\frac{m}{vp_1}.
\end{equation}

Putting together the above two equations gives
\begin{equation}
m=p_1(2p_1n+1)v^2\qquad l=(2p_1n+1)v.
\end{equation}

Again, choosing the opposite sign for $l$ would require a slightly modified charge vector in the analogue of (\ref{totalK}) but the anyon theory of the hierarchy states would be the same. The total filling fraction is
\begin{equation}
\nu_{p_2}=\frac{p_2}{p_1p_2+1},
\end{equation}
where we relabeled $p_2=2n$. The total quantum dimension after condensation is
\begin{equation}
\mathcal{D}_{\eC\boxtimes\eD/\mathcal{A}}=\sqrt{2(p_1p_2+1)}.
\end{equation}

The daughter states are abelian, generated by $\bar{\v b}^{-p_1}$, which has topological spin $e^{-\frac{2\pi i p_1}{2(p_1p_2+1)}}$ and charge $\frac{1}{p_1p_2+1}$.
\subsubsection{Explicit relation to Haldane/Halperin states}
We will now show that $\eC\boxtimes\eD/\mathcal{A}$ is indeed equivalent to the theory described by the Lagrangian density (\ref{kmatrixcond}). Specifically, we will show that condensing $\mathcal{A}$ (with some trivial fermions) in $\eC\boxtimes\eD$ stacked with invertible quantum spin Hall layers is related to (\ref{kmatrixcond}) stacked with invertible quantum spin Hall layers by a $GL(N,\mathbb{Z})$ transformation with determinant $\pm 1$. We will focus on the particular example $p_1=3$ and $n=1$ discussed in Sec.~\ref{sex1} (with $m=15$ and therefore $v=1$), but our argument applies more generally to all the hierarchy states described in the previous section. 

We describe the sMTC by an abelian Chern-Simons theory with Lagrangian density
\begin{equation}
\mathcal{L}=-\frac{1}{4\pi}K_{IJ}a_Ida_J+\frac{q_I}{2\pi }a_I dA,
\end{equation} 

where $\{a_I\}$ are emergent dynamical gauge fields, $q_I$ is the charge vector, and $A$ is a background $U(1)$ gauge field. $K_{IJ}$ describing $\eC\boxtimes \eD$ for $p_1=3$ and $n=1$ is given by
\begin{equation}
K=\begin{pmatrix} 3 & 0\\ 0 & 15\end{pmatrix}.
\end{equation}

The charge vector is given by $q=\begin{pmatrix} 1 & 1\end{pmatrix}^T$. Anyons of the theory are labeled by integer vectors $l$. In particular, the anyon we would like to condense is the charge neutral boson $l=\begin{pmatrix} 1 & -5\end{pmatrix}^T$. It has charge $q_l=q^TK^{-1}l=0$ and spin $\theta_l=\pi l^TK^{-1}l=2\pi$.

To condense $l$, we introduce a Higgs field $\alpha$ whose current couples to the anyon we want to condense. Specifically, we add a term $(a_1-5a_2)\frac{d\alpha}{2\pi}$ to the Lagrangian density, giving the $K$ matrix
\begin{equation}\label{kmatq}
K_H=\begin{pmatrix} 3 & 0 & 1\\ 0 & 15 & -5\\ 1 & -5 & 0\end{pmatrix},
\end{equation}
with charge vector $q=\begin{pmatrix} 1 & 1 & 0\end{pmatrix}^T$. There is an issue with condensing the Higgs boson in this theory, however. This is because the Higgs has topological spin $2\pi$ rather than $0$; this would result in a Chern Simons action for $\alpha$ after integrating out the dynamical gauge fields. To remedy this issue, we introduce trivial blocks to the $K$ matrix, that do not change the topological order or $U(1)$ symmetry fractionalization pattern. The resulting $K$ matrix and charge vector are given by
\begin{equation}\label{totalK}
\tilde{K}=\begin{pmatrix} 3 & 0 & 0 & 0 & 0 & 0 \\ 0 & 15 & 0 & 0 & 0 & 0\\ 0 & 0 & -1 & 0  & 0 & 0\\ 0 & 0 & 0 & -1 & 0 & 0\\ 0 & 0 & 0  & 0 & 1 & 0\\ 0 & 0 & 0  & 0 & 0 & 1\end{pmatrix}\qquad q=\begin{pmatrix} 1\\ 1 \\ 1 \\ 1 \\ 1 \\ 1 \end{pmatrix}.
\end{equation}

Now we can condense $l=\begin{pmatrix} 1 &  -5 & 1 & -1 & 0 & 0\end{pmatrix}^T$ which has topological spin and charge zero. Including the Higgs condensate, we have the $K$ matrix
\begin{equation}
\tilde{K}_H=\begin{pmatrix} \tilde{K} & l\\ l^T & 0\end{pmatrix}.
\end{equation}

$K$ matrices are equivalent under conjugation by a $GL(N,\mathbb{Z})$ transformation with determinant $\pm 1$. In this particular case, we can perform a $GL(7,\mathbb{Z})$ transformation given explicitly by
\begin{equation}\label{xmat}
X=\begin{pmatrix} 1 & -1 & 0 & 0 & 0 & 0 & 0 \\
  -1 & 0 & 0 & 0 & 0 & 0 & 0\\
  -3 & 0 & 1 & 1 & 0 & 0 & 0\\
  3 & 0 & -1 & 0 & 1 & 0 & 0\\
  0 & 0 & 0 & 0 & 0 & 1 & 0\\
  0 & 0 & 0 & 0 &0 & 0 & 1\\
  -3 & 0 & 1 & 0 & 0 & 0 & 0\end{pmatrix},
\end{equation}
which satisfies $X^T \tilde{K}_H X=K_{\mathrm{qp}}$ where
\begin{equation}\label{target}
K_{\mathrm{qp}}=\begin{pmatrix} 0 & 0 & 0 & 0 & 0 & 0 & 0 \\ 0 & 3& -1 & 0 & 0 & 0 & 0 \\ 0 & -1 & 2 & 0 & 0 & 0 & 0\\  0 & 0 & 0 & -1 & 0 & 0 & 0\\ 0 & 0 & 0 & 0 & -1 & 0 & 0\\ 0 & 0 & 0 & 0 &0 & 1 & 0\\ 0 & 0 & 0 & 0 & 0 & 0 & 1\end{pmatrix}
\end{equation}

This is simply the hierarchy state (\ref{kmatrixcond}) stacked with trivial blocks, together with a Higgs field with no Chern Simons term. Therefore, $\eC\boxtimes\eD$ (stacked with trivial theories) is equivalent to (\ref{kmatrixcond}) (stacked with trivial theories) after condensation of $\mathcal{A}$. In Appendix~\ref{sgln}, we explain how to derive $GL(N,\mathbb{Z})$ transformations like (\ref{xmat}).

\subsection{Pfaffian and anti-Pfaffian}\label{spfaff}

We will determine the daughter states for the Pfaffian state; those of the anti-Pfaffian state can be obtained by applying particle-hole conjugation to the Pfaffian daughter states. We will reproduce the results of \Rf{levin2009} using just the algebraic theory of anyons. 

The Pfaffian sMTC has anyons
\begin{equation}\label{pfaffanyons}
\{\v 0,\v a^2,\v a^4,\v a^6,\psi,\psi \v a^2,\psi \v a^4,\psi\v a^6,\sigma \v a,\sigma \v a^3,\sigma \v a^5,\sigma \v a^7\},
\end{equation}
where $\v a$ is the generator of the anyons in $U(1)_8$ and $\psi \v a^4$ is a transparent fermion. $\v 0,\psi,$ and $\sigma$ form an Ising MTC, with spins $1,-1,$ and $e^{2\pi i/16}$ respectively. We begin with the quasihole hierarchy, which produces topological orders with filling fraction slightly less than $1/2$. The minimal quasihole is $\sigma \v a$, so this is the anyon we would like to pair up with an anyon in the stacked theory $\eD$ to condense.

First, we choose a minimal modular extension. The simplest one is $\mathrm{Ising}\boxtimes U(1)_8$ (the other minimal modular extensions can be obtained by repeatedly stacking with $\mathrm{Ising}$ and condensing the pair of fermions $\psi \v a^4$ from the Pfaffian theory together with $\psi$ from $\mathrm{Ising}$). Next, we stack this theory with another (bosonic) topological order $\eD$ such that $\mathrm{Ising}\boxtimes U(1)_8\boxtimes\eD$ has a condensable algebra. As shown in Sec.~\ref{sconstraints}, an abelian MTC does not suffice. We therefore use $\overline{\mathrm{Ising}}\boxtimes U(1)_{-m}$ as an ansatz. Since $\eC^{(b)}\boxtimes \eD$ has an $\mathrm{Ising}\boxtimes\overline{\mathrm{Ising}}$ sector, it is natural to expect the condensable algebra to include terms like $\psi\bar{\psi}$ and $\sigma\bar{\sigma}$. Let us denote the generator of $U(1)_{-m}$ by $\bar{\v b}$. In order for $\sigma \v a\bar{\sigma} \bar{\v b}^l$ to be a boson, we must have

\begin{equation}\label{mpfaff}
\frac{l^2}{2m}=\frac{16n+1}{16}.
\end{equation}

For $\sigma \v a\bar{\sigma} \bar{\v b}^l$ to have charge zero, we require
\begin{equation}\label{lpfaff}
\frac{vl}{m}=\frac{1}{4}.
\end{equation}

Putting together (\ref{mpfaff}) and (\ref{lpfaff}) gives
\begin{equation}
m=2(16n+1)v^2\qquad l=\frac{1}{2}(16n+1)v.
\end{equation}

This means that $v$ must be even.\footnote{If in addition we wanted $\eD$ to be the minimal modular extension of a fermionic theory, then we would require $\bar{\psi}\bar{\v b}^{m/2}$ to carry odd integer charge. In this case, $v$ must be twice an odd integer.} The daughter states have filling fraction
\begin{equation}\label{pfaffhole}
\nu_{n}=\frac{1}{2}-\frac{1}{2(16n+1)}=\frac{8n}{16n+1},
\end{equation}
which matches with \Rf{levin2009}. 

We must check that $\sigma \v a\bar{\sigma} \bar{\v b}^l$ can be part of a condensable algebra. Fusing this anyon with itself gives
\begin{equation}
\sigma\bar{\sigma }\v a\bar{\v b}^{l}\times\sigma\bar{\sigma} \v a\bar{\v b}^{l}=\left(1+\psi+\bar{\psi}+\psi\bar{\psi}\right)\v a^2\bar{\v b}^{2l}.
\end{equation}

Unlike in (\ref{sigmafusion}), there are two bosons in the product, $\v a^2\bar{\v b}^{2l}$ and $\psi\bar{\psi}\v a^2\bar{\v b}^{2l}$. Therefore, (\ref{ninequ}) does not rule out $\sigma \v a\bar{\sigma} \bar{\v b}^l$ being in a condensable algebra. We find the following condensable algebra in $\eC^{(b)}\boxtimes \eD$: 
\begin{align}
\begin{split}
\mathcal{A}&=\v 0+\psi\bar{\psi}+\v a^2\bar{\v b}^{2l}+\v a^2\bar{\v b}^{2l}\psi\bar{\psi}\\
&+\dots+\v a^{4v-2}\bar{\v b}^{l(4v-2)}+\v a^{4v-2}\bar{\v b}^{l(4v-2)}\psi\bar{\psi}\\
&+\sigma\bar{\sigma}\v a\bar{\v b}^{l}+\sigma\bar{\sigma}\v a^3\bar{\v b}^{-3l}+\dots+\sigma\bar{\sigma}\v a^{4v-1}\bar{\v b}^{l(4v-1)}.
\end{split}
\end{align}

This condensable algebra has quantum dimension $8v$, while $\eC^{(b)}\boxtimes\eD$ has total quantum dimension $16v\sqrt{(16n+1)}$, so we find that the resulting topological order after condensation has total quantum dimension $2\sqrt{(16n+1)}$. This is an abelian MTC (non-abelian anyons like $\sigma\bar{\sigma}$ that survive the condensation get split).

After condensing the fermions, by removing anyons that braid nontrivially with $\psi \v a^4\sim\bar{\psi}\v b^{m/2}$ (where the equivalence is up to condensed bosons), we obtain an abelian fermionic phase with quantum dimension $\sqrt{2(16n+1)}$, generated by $\bar{\psi}\bar{\v b}^{2v}$. This anyon has topological spin $e^{(16n-1)\pi i/(16n+1)}$, which evaluates to $e^{15\pi i/17}$ for $n=1$, matching with \Rf{levin2009}. Its charge, computed from the braiding with $\v a^2\bar{\v b}^v$ (which survives the condensation), is $\frac{2v^2}{m}=\frac{1}{16n+1}$. The total chiral central charge, which is easily read off from the total chiral central charge of $\eC^{(b)}\boxtimes \eD$, is zero.

To check that the condensable algebra above describes a valid domain wall between $\mathrm{Ising}\times U(1)_8$ and $\overline{\mathrm{Ising}}\times U(1)_{-136}$, we observe that it gives the same result as sequentially condensing two abelian subgroups. First we condense $1+\psi\bar{\psi}$ to get $\mathbb{Z}_2$ gauge theory stacked with $U(1)_{8}\times U(1)_{-136}$, and then we condense the subgroup generated by $e \v a\bar{\v b}^l$ (which braids trivially with the transparent fermion) where $e$ is a boson in $\mathbb{Z}_2$ gauge theory. The first step divides the total quantum dimension by $2$ while the second step divides the total quantum dimension by $4v$. 

To produce the quasiparticle hierarchy, we instead stack $\eC^{(b)}$ with $\mathrm{Ising}\boxtimes U(1)_{m}$. We will differentiate the $\mathrm{Ising}$ factor in $\eC^{(b)}$ and that in $\eD$ by labeling the anyons of the latter by $\v 0,\psi',\sigma'$. The minimal quasiparticle in $\eC^{(b)}$ is $\sigma \v a^{-1}$. In order for $\sigma \v a^{-1}\sigma' \v b^l$ to be a boson, we need
\begin{equation}
\frac{l^2}{2m}=\frac{16n-3}{16}.
\end{equation}

For $\sigma\v a^{-1}\sigma'\v b^l$ to have charge zero, we require
\begin{equation}
\frac{vl}{m}=\frac{1}{4}.
\end{equation}

The above two equations give
\begin{equation}
m=2(16n-3)v^2\qquad l=\frac{1}{2}(16n-3)v.
\end{equation}

Again, $v$ must be even. The daughter states have filling fraction
\begin{equation}
\nu_{n}=\frac{1}{2}+\frac{1}{2(16n-3)}=\frac{8n-1}{16n-3}
\end{equation}

Again, we must check that $\sigma \v a^{-1}\sigma'\v b^l$ can be part of a condensable algebra. It fuses with itself to give
\begin{equation}
\sigma \sigma'\v a^{-1}\v b^{l}\times \sigma \sigma'\v a^{-1}\v b^{l}=(1+\psi+\psi'+\psi\psi')\v a^{-2}\v b^{2l}.
\end{equation}

There are two bosons, $\psi\v a^2\v b^{2l}$ and $\psi' \v a^2\v b^{2l}$, so there is no obstruction to $\sigma \v a^{-1}\sigma'\v b^{l}$ from (\ref{ninequ}). We propose the following condensable algebra in $\eC^{(b)}\boxtimes\eD$:
\begin{align}
\begin{split}
\mathcal{A}&=\v 0+\psi\psi'+\psi\v a^2\v b^{2l}+\psi'\v a^2\v b^{2l}+\v a^4\v b^{4l}\\
&+\dots+\v a^{4v-4}\v b^{l(4v-4)}+\psi\psi'\v a^{4v-4}\v b^{l(4v-4)}\\
&+\psi\v a^{4v-2}\v b^{l(4v-2)}+\psi'\v a^{4v-2}\v b^{l(4v-2)}+\sigma\sigma'\v a\v b^l\\
&+\dots + \sigma\sigma'\v a^{4v-1}\v b^{l(4v-1)}.
\end{split}
\end{align}

This condensable algebra has quantum dimension $8v$. $\eC^{(b)}\boxtimes\eD$ has total quantum dimension $16v\sqrt{(16n-3)}$, so the resulting MTC has total quantum dimension $2\sqrt{(16n-3)}$. Condensing the fermion gives an abelian sMTC with total quantum dimension $\sqrt{2(16n-3)}$, generated by an anyon $\psi' \v b^{2v}$ with topological spin $e^{(16n-1)\pi i/(16n-3)}$ and charge $\frac{1}{16n-3}$, recovering the results from \Rf{levin2009}. The chiral central charge of the daughter states are all 3; this can be read off from the chiral central charge of $\eC^{(b)}\boxtimes \eD$.

Similar to the condensation for the quasihole hierarchy, the above condensation is equivalent to two sequential abelian condensations. This allows us to justify the above condensation without computing $M$ symbols. The two-step abelian condensation proceeds as follows. First, we condense $\psi\psi'$, which takes $\mathrm{Ising}\times\mathrm{Ising}\to U(1)_4$. Then we condense the subgroup generated by $e \v a\v b^l$, where $e$ is the generator of $U(1)_4$. We condense $e \v a\v b^l$ because it braids trivially with the anyon we identify with the transparent fermion. 

As discussed in \Rf{levin2009}, the hierarchy states from the anti-Pfaffian are just obtained by particle-hole conjugation. Specifically, the quasihole hierarchy states have filling fraction
\begin{equation}
\nu_{n}=\frac{1}{2}+\frac{1}{2(16n+1)}=\frac{8n+1}{16n+1},
\end{equation}
while the quasiparticle hierarchy states have filling fraction
\begin{equation}
\nu_{n}=\frac{1}{2}-\frac{1}{2(16n-3)}=\frac{8n-2}{16n-3}.
\end{equation}

\subsubsection{Other choices for $\eD$}\label{sother}
We could have chosen other theories $\eD$ to stack onto $\eC^{(b)}$, that do not have Ising factors. For example, for the Pfaffian quasihole hierarchy, we can stack $\eC^{(b)}$ with $\mathrm{SU(2)}_{-2}\boxtimes U(1)_{-m}$. $SU(2)_{-2}$ is very similar to $\overline{\mathrm{Ising}}$, and we discuss its hierarchy states in Sec.~\ref{ssu22}. The difference between the two MTCs is that the Ising-like anyon $\bar{\mathbf{\frac{1}{2}}}$ in $SU(2)_{-2}$ has spin $e^{-2\pi i3/16}$ rather than $e^{-2\pi i/16}$ (concretely, we can get $SU(2)_{-2}$ by stacking $\overline{\mathrm{Ising}}$ with $U(1)_{-4}$ and condensing a diagonal boson). Suppose that we tried to construct a condensable algebra that contains $\sigma \v a\bar{\mathbf{\frac{1}{2}}}\bar{\v b}^{l}$ then we must have
\begin{equation}
\frac{l^2}{2m}=\frac{16n-1}{16}.
\end{equation}

Together with the zero charge condition $l^2=\frac{m^2}{16v^2}$, we have
\begin{equation}
m=2(16n-1)v^2\qquad l=\frac{1}{2}(16n-1)v.
\end{equation}

The resulting hierarchy states would have filling fraction
\begin{equation}
\nu_{n}=\frac{1}{2}-\frac{1}{2(16n-1)}=\frac{8n-1}{16n-1}.
\end{equation}

We can check that fusing $\sigma a\bar{\mathbf{\frac{1}{2}}}\bar{b}^{l}$ with itself gives
\begin{equation}
\sigma \v a\bar{\mathbf{\frac{1}{2}}}\bar{\v b}^{l}\times \sigma \v a\bar{\mathbf{\frac{1}{2}}}\bar{\v b}^{l}=(\mathbf{0}+\psi+\overline{\mathbf{1}}+\psi\overline{\mathbf{1}})\v a^2\bar{\v b}^{2l}.
\end{equation}

Two of the anyons in the fusion product, $\psi \v a^2\bar{\v b}^{2l}$ and $\bar{\mathbf{1}} \v a^2\bar{\v b}^{2l}$ are bosons (recall that $\bar{\v 1}$ has spin $-1$). Continuing to fuse in $\psi \v a^2\bar{\v b}^{2l}$, we find that there are always bosons in the fusion product, so there is no obstruction from (\ref{ninequ}). We propose the condensable algebra for $v=2$ (the generalization to other $v$ is straightforward):
\begin{align}
\begin{split}
\mathcal{A}&=\mathbf{0}+\psi\bar{\mathbf{1}}+\psi \v a^2\bar{\v b}^{2l}+\bar{\mathbf{1}}\v a^2\bar{\v b}^{2l}+\v a^4\bar{\v b}^{4l}\\
&+\psi\bar{\mathbf{1}}\v a^4\bar{\v b}^{4l}+\psi\v a^6\bar{\v b}^{6l}+\bar{\mathbf{1}}\v a^6\bar{\v b}^{6l}+\sigma\bar{\mathbf{\frac{1}{2}}}\v a\bar{\v b}^{l}\\
&+\sigma\bar{\mathbf{\frac{1}{2}}} \v a^3\bar{\v b}^{3l}+\sigma\bar{\mathbf{\frac{1}{2}}} \v a^5\bar{\v b}^{5l}+\sigma\bar{\mathbf{\frac{1}{2}}}\v a^7\bar{\v b}^{7l}.
\end{split}
\end{align}

The resulting theory is generated by $\bar{\v 1}\bar{\v b}^{2v}$, which has spin $e^{\pi i(16n-3)/(16n+1)}$ and charge $\frac{1}{16n-1}$. It is abelian theory with total quantum dimension $\sqrt{2(16n-1)}$. The quasiparticle hierarchy with $\eD=SU(2)_{2}\boxtimes U(1)_m$ would give
\begin{equation}\label{su2ising}
    \nu_n=\frac{8n-2}{16n-5}.
\end{equation}

These daughter states are actually a \emph{simpler} than those given in Sec.~\ref{spfaff}, because they have smaller rank. However, because the change in chiral central charge is larger ($SU(2)_{-2}\boxtimes U(1)_{-m}$ has chiral central charge $-\frac{5}{2}$ while $\overline{\mathrm{Ising}}\boxtimes U(1)_{-m}$ has chiral central charge $-\frac{3}{2}$) and the change in filling is larger, it may be less favorable in a physical setting. Specifically, the fact that one must tune through the filling fractions of the daughter states in Sec.~\ref{spfaff} before arriving at these fillings may make these daughter states less likely to be realized.

\subsection{Bosonic Pfaffian}\label{sbosonicpfaff}
Here we consider the bosonic Pfaffian state, which is given by a wavefunction similar to (\ref{Pfwv}) except with $\chi_1^2(\{z_i\})$ replaced by $\chi_1^p(\{z_i\})$, where $p$ is odd. This makes the wavefunction as a whole symmetric with respect to swapping two coordinates. We present these filling fractions because the filling fractions of the hierarchy states are simpler and the states themselves have lower rank. Therefore, these hierarchies may be easier to study numerically compared to the hierarchies for the Pfaffian state described above. \footnote{We thank D. Son for pointing this out.}

From the CFT construction for the wavefunction, one can show that the wavefunction with $p=1$ corresponds to an anyon theory with the $U(1)_{8}$ used in the fermionic Pfaffian by $U(1)_4$. Specifically, in $\mathrm{Ising}\times U(1)_4$, we take the physical boson to be $\psi \v a^2$ which has charge 1. The vison is then $\v a^2$. Condensing the physical boson gives the theory
\begin{equation}
    \{\mathbf{0},\v a^2,\sigma \v a\},
\end{equation}
which matches with the $SU(2)_2$ MTC, aligned with Ref.~\cite{fradkin1998}. The minimal quasihole is $\sigma \v a$, which has charge $1$ from braiding with the vison $\v a^2$.

To construct the quasihole hierarchy, we stack with $\overline{\mathrm{Ising}}\times U(1)_{-m}$, where $U(1)_{-m}$ is generated by $\bar{\v b}$. To condense the minimal quasihole $\sigma \v a$ fused with $\overline{\sigma}\bar{\v b}^l$, we find that
\begin{equation}
    m=(8n+1)v^2\qquad l=\frac{1}{2}(8n+1)v.
\end{equation}

This gives states at filling fraction
\begin{equation}
    \nu_n=\frac{8n}{8n+1}.
\end{equation}

The simplest state occurs at filling fraction $\nu_1=\frac{8}{9}$, with chiral central charge zero. Note that this filling fraction matches that of the bosonic Jain state $\frac{p}{p+1}$ with $p=8$, similar to how the fermionic Pfaffian quashihole hierarchy matches with the fermionic Jain state at filling fraction $\frac{p}{2p+1}$ with $p=8$.\cite{cooper2001} Like in the fermion case, the state differs from the Jain state at the same filling by chiral central charge 8.

For the quasiparticle hierarchy, we stack with $\mathrm{Ising}\times U(1)_{m}$ where $U(1)_m$ is generated by $\v b$. To condense the minimal quasiparticle $\sigma\v a$ we get
\begin{equation}
    m=2(4n-1)v^2\qquad l=(4n-1)v,
\end{equation}
which gives states at filling fraction
\begin{equation}
    \nu_n=\frac{8n-1}{8n-2}
\end{equation}

The simplest state occurs at $\nu_1=\frac{7}{6}$, with chiral central charge 3. Again, similar to the fermionic case, the filling fraction matches that of a Jain state but with chiral central charge offset by 8.

\subsection{PH-Pfaffian}
The particle-hole (PH)-Pfaffian theory has anyons
\begin{equation}
\{\mathbf{0},\v a^2,\v a^4,\v a^6,\bar{\psi},\bar{\psi}\v a^2,\bar{\psi}\v a^4,\bar{\psi}\v a^6,\bar{\sigma}\v a,\bar{\sigma}\v a^3,\bar{\sigma}\v a^5,\bar{\sigma}\v a^7\},
\end{equation}
where $\v a$ is the generator of the anyons in $U(1)_{8}$. The anyon theory above can also be written as $\mathrm{Ising}\boxtimes U(1)_{-8}/\mathbb{Z}_2$. The simplest minimal modular extension is simply $\overline{\mathrm{Ising}}\boxtimes U(1)_{8}$. The minimal quasihole is $\bar{\sigma}\v a$. 

Like with the Pfaffian state, we can stack this theory with $\overline{\mathrm{Ising}}\boxtimes U(1)_{-m}$ for the quasihole hierarchy. $\eD$ in this case can give fully (anti)-holomorphic wavefunctions using correlation functions in those antichiral CFTs. We can also consider stacking with the time-reversal of the PH-Pfaffian, but with a different $U(1)$ factor, for the quasihole hierarchy. This naively would be unfavorable, because it would not allow us to use the CFT ansatz to obtain a antiholomorphic wavefunction for the excitations in $\eD$ (see Appendix~\ref{swavefunction}), but in this case the parent state itself already does not have an obvious holomorphic wavefunction.

If we stack with $\overline{\mathrm{Ising}}\boxtimes U(1)_{-m}$, to obtain a charge zero boson $\bar{\sigma}\v a\bar{\sigma}'\bar{\v b}^l$, we get the solutions
\begin{equation}\label{solph}
    m=2(16n-1)v^2\qquad l=\frac{1}{2}(16n-1)v,
\end{equation}
which gives filling fractions
\begin{equation}\label{quasiholeph}
    \nu_n=\frac{8n-1}{16n-1}.
\end{equation}

We can generate a condensable algebra from $\bar{\sigma}\v a\bar{\sigma}'\bar{\v b}^l$; two of the anyons in the fusion product of this anyon with itself are bosons. These are $\v a^2\bar{\v b}^{2l}$ and $\bar{\psi}\bar{\psi}'\v a^2\bar{\v b}^{2l}$. The hierarchy states are abelian, generated by $\bar{\psi}'\bar{\v b}^{2v}$, which has charge $-\frac{1}{16n-1}$ from braiding with $\v a^2\bar{\v b}^v$. They have chiral central charge -2.

To obtain the quasiparticle hierarchy, we stack with $\mathrm{Ising}\boxtimes U(1)_m$. The solutions are identical to those written in (\ref{solph}), resulting in hierarchy states at filling fractions
\begin{equation}
    \nu_n=\frac{8n}{16n-1}.
\end{equation}

These states are simply the particle-hole conjugates of the quasihole hierarchy states (\ref{quasiholeph}). While the quasihole and quasiparticle hierarchies of the Pfaffian state are very asymmetric, those of the PH-Pfaffian are simply related by particle-hole conjugation.

If we stack with $\mathrm{Ising}\boxtimes U(1)_{-m}$, we would obtain a different quasihole hierarchy. We would obtain hierarchy states at filling fractions
\begin{equation}\label{phpfaffhole}
    \nu_n=\frac{8n}{16n+1}.
\end{equation}

These states are identical to those in the quasihole hierarchy of the Pfaffian state (\ref{pfaffhole}). Similarly, stacking with $\overline{\mathrm{Ising}}\boxtimes U(1)_m$ for the quasiparticle hierarchy would give
\begin{equation}
    \nu_n=\frac{8n+1}{16n+1}.
\end{equation}

These states are the particle-hole conjugates of those in (\ref{phpfaffhole}). The chiral central charge for both the quasihole and the quasiparticle hierarchy states is zero. This makes them distinct from Jain states at the same filling.
\subsection{$\chi_1\chi_2^2$ FQH state}\label{ssu22}

We can also study the hierarchy states for the $SU(2)_2$ FQH state\cite{wen1991}
\begin{equation}
\Psi(\{z_i\})=\chi_{1}(\{z_i\})[\chi_{2}(\{z_i\})]^2.
\end{equation}

This state is non-abelian since it is described by $SU(2)_2$ Chern-Simons effective theory. 
The anyons in such a state are described by the $U(1)^4/SU(2)_2 = SU(2)_2\boxtimes U(1)_8/\mathbb{Z}_2$ sMTC, which is a slight modification of the Pfaffian theory. The anyon content is the same as in (\ref{pfaffanyons}), except the $\sigma$ has topological spin $e^{3\pi i /16}$. If we use as $\eD$ $SU(2)_{-2}\boxtimes U(1)_{-m}$, then the quashihole daughter states are the same as for Pfaffian (\ref{pfaffhole}). We can also construct a quasiparticle hierarchy by stacking with $SU(2)_2\boxtimes U(1)_{m}$. We obtain $m=2(16n-7)v^2$ and $l=\frac{1}{2}(16n-7)v$, giving hierarchy states at filling fraction
\begin{equation}
\nu_n=\frac{8n-3}{16n-7}.
\end{equation}

These states are all abelian. For example, for $n=1$ the filling fraction is $\nu'=\frac{5}{9}$, and the abelian state is generated by an anyon with spin $e^{2\pi i /9}$.

If we use instead $\eD=\overline{\mathrm{Ising}}\boxtimes U(1)_{-m}$ for the quasihole hierarchy instead of $SU(2)_{-2}\boxtimes U(1)_{-m}$, we would get a different hierarchy. This hierarchy might be more favorable, because the chiral central charge of $\eD$ is smaller. In this case, we obtain
\begin{equation}
\nu_n=\frac{8n+1}{16n+3}.
\end{equation}

For the quasiparticle hierarchy with $\eD=\mathrm{Ising}\boxtimes U(1)_m$, we would get
\begin{equation}
\nu_n=\frac{8n-2}{16n-5},
\end{equation} 
which matches with (\ref{su2ising}). 

In Appendix~\ref{s16fold}, we generalize the above sections and compute the filling fractions for the hierarchy states for FQH phases at filling fraction $\frac{1}{2q}$ obtained from any theory in the 16-fold way together with $U(1)_{8q}$. For the non-abelian FQH states, we will always stack with $\mathrm{Ising}\times U(1)_{8q}$ or $\overline{\mathrm{Ising}}\times U(1)_{-8q}$ because this minimizes the change in chiral central charge between the parent state and the hierachy states.

\subsection{$\chi_2^3$ FQH state}\label{ssu23}
We now consider the $SU(2)_3$ FQH state, which occurs at filling fraction $\nu=\frac{2}{3}$ and is given by the wavefunction\cite{wen1991} 
\begin{equation}
\Psi(\{z_i\})=[\chi_{2}(\{z_i\})]^3.
\end{equation}

It was argued in Refs.~\cite{wen1991,blok1992} that since the electrons in this state lie in the first four Landau levels (and more generally, for $[\chi_2(\{z_i\})]^k$ the electrons lie in the first $k+1$ Landau levels), the wavefunction is not holomorphic. However, we can treat the Landau level as a flavor index and instead consider electrons in the lowest Landau level carrying four flavors from a spin $\frac{3}{2}$ representation of $SU(2)$. The above state should therefore be described by $SU(2)_3$ MTC together with a $U(1)$ factor, \ie by $ SU(2)_3\boxtimes U(1)_6/\mathbb{Z}_2$.

In Appendix~\ref{ssu22}, we derive the quasiparticles and filling of this state using the CFT ansatz for the wavefunction. Here we summarize the results: the theory has anyons given by the $SU(2)_3\boxtimes U(1)_6/\mathbb{Z}_2$ sMTC
\begin{equation}\label{su23anyons}
\left\{\mathbf{0},\v a^2,\v a^4,\mathbf{\frac{1}{2}}\v a,\mathbf{\frac{1}{2}}\v a^3,\mathbf{\frac{1}{2}}\v a^5,\mathbf{\frac{3}{2}}\v a,\mathbf{\frac{3}{2}}\v a^3,\mathbf{\frac{3}{2}}\v a^5,\mathbf{1}\v,\mathbf{1}\v a^2,\mathbf{1}\v a^4\right\},
\end{equation}
where $\{\mathbf{0}, \mathbf{\frac{1}{2}},\mathbf{1},\mathbf{\frac{3}{2}}\}$ are elements of the $SU(2)_3$ MTC with topological spins $\{1,e^{2\pi i3/20},e^{2\pi i2/5},e^{2\pi i3/4}\}$, and $\v a$ generates $U(1)_6$. Using the spins and the $SU(2)_3$ fusion rules, it is straightforward to check that $\mathbf{\frac{3}{2}}\v a^3$ is indeed a fermion that braids trivially with all the other anyons listed above. $\v v=\v a^2$ to give this transparent fermion odd integer charge. From the spin of $\v a^2$ we recover the fact that this theory occurs at filling fraction $\nu=\frac{2}{3}$. Note that the $SU(2)_3$ MTC can be written as $(G_2)_1\boxtimes U(1)_{-2}$, where $(G_2)_1$ gives the Fibonacci anyon theory. The $\mathbf{1}$ comes from $(G_2)_1$ while $\mathbf{\frac{3}{2}}$ generates $U(1)_{-2}$. Recall that the chiral central charge of the Fibonacci anyon theory, $(G_2)_1$ is $\frac{14}5$ and its quantum dimension is ${\mathcal D} = \sqrt{1+\phi^2}$, where $\phi=(1+\sqrt 5)/2$ is the golden ratio. The minimal modular extension is simply $SU(2)_3\boxtimes U(1)_6$. 

Unlike the previous FQH phases, this FQH phase has \emph{two} different minimal quasiholes (and quasiparticles). The two minimal quasiholes are $\mathbf{\frac{1}{2}}\v a$ and $\mathbf{\frac{3}{2}}\v a$. These anyons both have with charge $\frac{1}{3}$ from braiding with $\v a^2$. Therefore, we can construct two sets of hierarchies: one from condensing only abelian anyons, generated by the minimal abelian quasihole, and one from condensing the non-abelian minimal quasihole. We will find that the abelian minimal quasihole gives hierarchy states that are non-abelian, but have smaller rank, while the non-abelian minimal quasiholes give abelian hierarchy states that have much larger rank.

We begin with the abelian condensation. Because we are only condensing abelian anyons, we do not need to use condensable algebras and minimal modular extension. We choose $\eD$ to be the simplest theory that has anyons that can bind with $\mathbf{\frac{3}{2}}\v a$ to form a condensable boson, which is $U(1)_{-m}$. Requiring $\mathbf{\frac{3}{2}}\v a\bar{\v b}^l$ to be a charge zero boson gives
\begin{equation}
    m=3(6n+5)v^2\qquad l=(6n+5)v.
\end{equation}

This results in hierarchy states with filling fraction
\begin{equation}
    \nu_n=\frac{4n+3}{6n+5}.
\end{equation}

These states are non-abelian. For example, $\mathbf{\frac{1}{2}}\v a\bar{\v b}^{2v}$ braids trivially with the condensed anyons, and survives the condensation. Furthermore, these hierarchy states have chiral central charge $\frac{9}{5}$. This also indicates that they are non-abelian, because abelian states can only have integer chiral central charge. Yet another simple way to show that these states are non-abelian is by using (\ref{dimchange}), which says that the total quantum dimension of the resulting theory is given by the total quantum dimension of the original theory divided by the quantum dimension of the condensable algebra (or in this case, because we can use condensable subgroups, the number of anyons in the subgroup). The dimension of the condensable algebra in this case must be integer because it consists only of abelian anyons. A condensable algebra with integer quantum dimension cannot remove the golden ratios (coming from the quantum dimensions of $\mathbf{\frac{1}{2}}$ and $\mathbf{1}$) in the numerator coming from the total quantum dimension of $\eC\boxtimes\eD$, so the hierarchy states must also have total quantum dimension that includes golden ratios. Specifically, an abelian topological order cannot have total quantum dimension $\sqrt{2(1+\phi^2)(6n+5)}$ like these hierarchy states, where $\phi$ is the golden ratio. 

For the quasiparticle hierarchy, we stack with $U(1)_m$ and get 
\begin{equation}
    m=3(6n-5)v^2\qquad l=(6n-5)v.
\end{equation}

The hierarchy states have filling fraction 
\begin{equation}\label{su23ab}
    \nu_n=\frac{4n-3}{6n-5}.
\end{equation}

Again, this theory is non-abelian, with chiral central charge $\frac{19}{5}$.

Now, let us turn to the other hierarchy obtained from non-abelian anyon condensation. In contrast to earlier examples that either involved abelian anyon condensation, or even the previous non-abelian anyon condensation example that could be replicated by a two stage abelian condensation, here we will be faced with a technically more challenging situation where such simplifications are absent. We construct a different hierarchy from condensing non-abelian anyons $\mathbf{\frac{1}{2}}\v a$. In this case, we stack with $SU(2)_{-3}\boxtimes U(1)_{-m}$ to obtain the anyon $\mathbf{\frac{1}{2}}\v a\overline{\mathbf{\frac{1}{2}}}\bar{\v b}^l$. For this anyon to be a charge zero boson, we must have
\begin{equation}
    m=\frac{3}{2}(12n+1)v^2\qquad l=\frac{1}{2}(12n+1)v.
\end{equation}

The filling fraction of these daughter states are
\begin{equation}
\nu_{n}=\frac{8n}{12n+1},
\end{equation}
and they have zero chiral central charge.

For $v=2$, we propose the following condensable algebra:
\begin{align}
\begin{split}
\mathcal{A}&=\mathbf{0}+\v a^2\bar{\v b}^{2l}+\v a^4\bar{\v b}^{4l}+\mathbf{\frac{3}{2}}\bar{\mathbf{\frac{3}{2}}}\v a \bar{\v b}^{l}+\mathbf{\frac{3}{2}}\bar{\mathbf{\frac{3}{2}}}\v a^3\bar{\v b}^{3l}\\
&+\mathbf{\frac{3}{2}}\bar{\mathbf{\frac{3}{2}}}\v a^5\bar{\v b}^{5l}+\mathbf{1}\bar{\mathbf{1}}+\mathbf{1}\bar{\mathbf{1}}\v a^2\bar{\v b}^{2l}+\mathbf{1}\bar{\mathbf{1}}\v a^4\bar{\v b}^{4l}\\
&+\mathbf{\frac{1}{2}}\bar{\mathbf{\frac{1}{2}}}\v a\bar{\v b}^{l}+\mathbf{\frac{1}{2}}\bar{\mathbf{\frac{1}{2}}}\v a^3\bar{\v b}^{3l}+\mathbf{\frac{1}{2}}\bar{\mathbf{\frac{1}{2}}}\v a^5\bar{\v b}^{5l}.
\end{split}
\end{align}

The total quantum dimension of $\eC^{(b)}\boxtimes\eD$ is $12(1+\phi^2)\sqrt{12n+1}$ where $\phi$ is the golden ratio, and the condensable algebra has dimension $6(1+\phi^2)$. Therefore, the resulting theory has total quantum dimension $2\sqrt{12n+1}$. Condensing the fermion gives an abelian sMTC generated by $\bar{\mathbf{\frac{3}{2}}}\bar{\v b}^3$ with spin $e^{2\pi i/4}e^{2\pi i3^2/(2m)}=e^{(6n-1)\pi i/(12n+1)}$. Note that these abelian topological orders are distinct from states in the Jain sequence, as can be seen from the filling fraction.

Now we turn to the quasiparticle hierarchy from condensation of the non-abelian anyon $\mathrm{\frac{1}{2}}\v a^5$. To condense this anyon, we can try to stack with $SU(2)_{3}\boxtimes U(1)_m$. In order for $\mathbf{\frac{1}{2}}\v a^5\mathbf{\frac{1}{2}}\v b^l$ to be a charge zero boson, we must have
\begin{equation}
m=\frac{3}{10}(60n-23)v^2\qquad l=\frac{1}{10}(60n-23)v
\end{equation}

However, with these solutions for $l$ and $m$, fusing $\mathbf{\frac{1}{2}}\v a^5\mathbf{\frac{1}{2}}\v b^l$ with itself does not produce any bosons. Therefore, by (\ref{ninequ}), this does not produce a valid condensable algebra. 

Instead, we can use a FQH state based on the $SU(3)_2$ MTC. This is likely to give a condensable algebra because 
\begin{equation}
    SU(2)_{-3}\leftrightarrow SU(3)_2\times U(1)_6/\mathbb{Z}_2,
\end{equation}
according to level-rank duality (up to transparent fermions)\cite{hsin2016}. The $SU(2)_{-3}$ theory is desirable because it gives a condensable algebra when stacked with $SU(2)_3$. $SU(3)_2$ is actually the simplest theory (with smallest chiral central charge, that does not have additional non-abelian anyons besides those with Fibonacci and abelian fusion rules) that contains anyons that are the time-reversal (with opposite spin) of the Fibonacci anyon in $SU(2)_3$.

The $SU(3)_2$ MTC has primary fields $\{\mathbf{0},\mathbf{3},\overline{\mathbf{3}},\mathbf{8},\mathbf{6},\overline{\mathbf{6}}\}$ where $\{\mathbf{0},\mathbf{6},\overline{\mathbf{6}}\}$ form an abelian $\mathbb{Z}_3$ theory with spins $\{1,e^{2\pi i 2/3},e^{2\pi i 2/3}\}$. $\mathbf{8}$ has spin $\frac{3}{5}$, which is the opposite of that of $\mathbf{1}$ in $SU(2)_3$; $\mathbf{3}$ and $\overline{\mathbf{3}}$ are obtained by fusion of $\mathbf{8}$ with $\overline{\mathbf{6}}$ and $\mathbf{6}$ respectively (for detailed fusion rules, see i.e. Ref.~\cite{cordova2023}). The chiral central charge is $\frac{16}{5}$.

We can try to obtain a condensable boson by fusing $\mathbf{\frac{1}{2}}\v a^{-1}$ with one of the (anti) Fibonacci-like anyons $\{\mathbf{3},\overline{\mathbf{3}},\mathbf{8}\}$, together with $\v b^l$ from a $U(1)_m$ theory. If we condense $\mathbf{\frac{1}{2}}\v a^{-1}\mathbf{8} \v b^l$, then we get
\begin{equation}
    m=3(6n-5)v^2\qquad l=(6n-5)v,
\end{equation}
so the hierarchy states occur at filling fraction
\begin{equation}
    \nu_n=\frac{4n-3}{6n-5},
\end{equation}
which match with (\ref{su23ab}) despite the fact that these states are abelian. Let us consider $n=1$ in more detail.\footnote{While this seems to occur at $\nu=1$, the parent state is predicted in a setting with four degenerate Landau levels\cite{TW230809702}, so the filling is actually $\frac{1}{4}$ after including the degeneracy.} For $n=1$, we choose $v=2$ for $\eD$ to be bosonic. The condensable algebra is given by
\begin{equation}
    \mathcal{A}=\mathbf{0}+\mathbf{1}\times\mathbf{8}+\mathbf{\frac{3}{2}}\v a^{-1}\v b^{2}+\mathbf{\frac{1}{2}}\v a^{-1}\mathbf{8}\v b^{2}+\cdots
\end{equation}
The resulting theory is abelian, with total quantum dimension $\sqrt{36}$, generated by $\mathbf{\frac{3}{2}}\v b$ and $\mathbf{6}$, with chiral central charge $\frac{9}{5}+1+\frac{16}{5}+1=7$. Condensing the fermion $\v b^6\sim \mathbf{\frac{3}{2}}\v a^3$ results in an abelian theory generated by $\v b^2$ and $\mathbf{6}$, which has the anyon content of $U(1)_3$ Chern-Simons theory stacked with cyclic $\mathbb{Z}_3$ anyons generated by $\mathbf{6}$. As mentioned in the next section, the latter theory can be obtained from $U(1)_{-2}\times U(1)_{-6}$ by condensing an order two anyon. 

We can also consider generating a condensable algebra from $\mathbf{\frac{1}{2}}\v a^{-1}\mathbf{3}\v b^l$. For this anyon to be a charge zero boson, we require that
\begin{equation}
    m=9(2n-1)v^2\qquad l=3(2n-1)v.
\end{equation}

The hierarchy states have filling fraction
\begin{equation}
    \nu_n=\frac{1}{9}\left(\frac{12n-5}{2n-1}\right).
\end{equation}

For $v=2$ and $n=1$, we get $\nu=\frac{7}{9}$, with $m=32,l=6$. These are higher rank than those described above, so are less likely to be stable.

\subsection{$\mathbb{Z}_3$ Read-Rezayi ($\mathrm{Pf}_3$)}\label{sz3}
Let us now turn to the $\mathbb{Z}_3$ Read-Rezayi state \cite{ReadRezayi1,ReadRezayi2}, which occurs at filling fraction $\frac{3}{5}$ and, like the previous example, hosts Fibonacci anyons. Moreover, the observation \cite{xia2004,pan2008,choi2008,kumar2010,zhang2012} of a quantum Hall plateau at $2+\frac25 $, albeit with a small gap, combined with numerical simulations that incorporate Landau level mixing \cite{wojs2009,zhu2015,mong2017,pakrouski2016}, indicate that experiments realize the particle-hole conjugate of the $\mathbb{Z}_3$ Read Rezayi state. It is therefore of significant interest to utilize the tools developed here to predict a new hierarchy of potential daughter states. Our calculations below, to compute these daughter states, will parallel those of the previous section.\footnote{Note that the Read-Rezayi state belongs in a family of states labeled by $k$, at filling fraction $\frac{k}{k+2}$. We discuss the $k=3$ state in this section. The $k=2$ state is simply the Pfaffian state described in Sec.~\ref{spfaff}. The more general sequence of Read-Rezayi states is summarized nicely in Ref.~\cite{bonderson2012}.}

The $\mathbb{Z}_3$ Read-Rezayi state is based on the $\mathbb{Z}_3$ parafermion CFT, which has primary fields $\{\psi_1,\psi_2,\sigma_1,\sigma_2,\epsilon\}$ with topological spins $\{e^{2\pi i 2/3},e^{2\pi i 2/3},e^{2\pi i /15},e^{2\pi i /15},e^{2\pi i 2/5}\}$. The fusion rules are given by cyclic $\mathbb{Z}_3$ fusion generated by $\psi_1$ (with $\psi_1\times\psi_1=\psi_2$ and $\psi_2\times\psi_1=\mathbf{0}$)\footnote{This theory can be obtained by $U(1)_{-2}\times U(1)_{-6}$ after condensing the diagonal order two anyon. From this, it is easy to see that the chiral central charge of the $\mathbb{Z}_3$ Read Rezayi theory is $\frac{14}{5}-2=\frac{4}{5}$.} together with Fib ($\epsilon\times\epsilon=1+\epsilon$). Fusing any of the $\mathbb{Z}_3$ fields with $\epsilon$ gives another $\epsilon$-like field:
\begin{equation}
\psi_1\times\epsilon=\sigma_2\qquad \psi_2\times\epsilon=\sigma_1,
\end{equation}
with
\begin{equation}
\sigma_2\times\sigma_2=\psi_2+\sigma_1\qquad \sigma_1\times\sigma_1=\psi_1+\sigma_2.
\end{equation}

The anyon theory is given by 
\begin{equation}\label{Z3rr}
\{\v a^{3j},\psi_1 \v a^{3j+2},\psi_2\v a^{3j+1},\sigma_1\v a^{3j+1},\sigma_2\v a^{3j+2},\epsilon \v a^{3j}\},
\end{equation}
where $\v a$ is the generator of $U(1)_{15}$, and the transparent fermion is $\psi_1\v a^5$. We abbreviated the above notation by writing $\{\v a^{3j}\}$ instead of $\{\mathbf{0},\v a^3,\v a^6,\dots,\v a^{12}\}$. $\psi_2\v a^{10}$ and $\psi_1 \v a^{20}$ are transparent bosons that we condense. \footnote{After condensing the transparent bosons, one can also write this as $\mathrm{Fib}$ with a minimal abelian theory labeled by $p=7,N=5$ according to the notation of Ref.~\cite{hsin2019}. This theory has 10 non-abelian anyons with Fibonacci fusion rules and 10 abelian anyons.} The filling fraction is $\nu=\frac{3}{5}$ according to the topological spin of $\v a^3$.  

Like in the previous section, there are actually two distinct minimal quasiholes and quasiparticles: $\sigma_1\v a$ and $\psi_2\v a$ are both minimal quasiholes (with charge $\frac{1}{5}$) and $\sigma_2 \v a^{-1}$ and $\psi_1\v a^{-1}$ are both minimal quasiparticles (with charge $-\frac{1}{5}$). Again, we find that condensing the abelian quasiholes and quasiparticles give non-abelian theories with smaller rank, and are therefore more favorable.

For the quasihole hierarchy obtained by condensing abelian quasiholes, we stack with $U(1)_{-m}$. In order for $\psi_2\v a\bar{\v b}^{l}$ to be a charge zero boson, we must have 
\begin{equation}
    m=5(10n+7)v^2\qquad l=(10n+7)v.
\end{equation}

 It follows that the hierarchy states have filling fraction
\begin{equation}
\nu_n=\frac{6n+4}{10n+7},
\end{equation}
with chiral central charge $\frac{4}{5}$.

For the quasiparticle hierarchy, we stack with $U(1)_m$ and get 
\begin{equation}
    m=5(10n-7)v^2\qquad l=(10n-7)v,
\end{equation}
so the hierarchy states have filling fraction
\begin{equation}\label{z3ab}
\nu_n=\frac{6n-4}{10n-7},
\end{equation}
and chiral central charge $\frac{14}{5}$. This gives another theory at the same filling fraction as the $SU(2)_3$ theory discussed in Sec.~\ref{ssu23} with Fibonacci anyons.

Both of these hierarchies are non-abelian. Note that for $n=1$, the (particle-hole conjugate of the) quasiparticle hierarchy gives a non-abelian state with Fibonacci anyons at filling fraction $\frac{1}{3}$. This is different from the non-abelian Pfaffian-like state obtained by Ref.~\cite{bonderson2008} (see Sec.~\ref{sbondsling}), because it carries anyons that have Fibonacci fusion rules.

We obtain a different set of hierarchy states if we instead choose to condense the non-abelian minimal quasihole and quasiparticle. For the quasihole hierarchy, we condense $\sigma_1 \v a$ by choosing $\eD$ to be simply the time reversal of (\ref{Z3rr}), except with a different $U(1)$ factor. Again we call $\bar{\v b}^l$ the anyons of $U(1)_{-m}$. To make $\sigma_1\v a\bar{\sigma}_1\bar{\v b}^l$ a charge zero boson, we must have
\begin{equation}
m=\frac{5}{3}(30n+1)v^2\qquad l=\frac{1}{3}(30n+1)v.
\end{equation}

Therefore, $v$ must be a multiple of three. The filling fractions of the hierarchy states are 
\begin{equation}
\nu_n=\frac{18n}{30n+1}.
\end{equation}

The condensable algebra can be generated by $\sigma_1\v a\bar{\sigma}_1\bar{\v b}^l$, and consists of anyons of the form $\sigma_1\v a^{(3j+1)l}\bar{\sigma}_1\bar{\v b}^{(3j+1)l}$, $\psi_2\v a^{(3j+1)l}\bar{\psi}_2\bar{\v b}^{(3j+1)l}$, $\sigma_2\v a^{(3j+2)l}\bar{\sigma}_2\bar{\v b}^{(3j+2)l}$, etc.

For the quasiparticle hierarchy, we use a similar theory as (\ref{Z3rr}) except with a different $U(1)$ factor. For $\sigma_2\v a^{-1}\sigma_1'\v b^l$ to be a charge zero boson, we have
\begin{equation}
    m=\frac{25}{3}(6n-1)v^2\qquad l=\frac{5}{3}(6n-1)v.
\end{equation}

However, we encounter the same issue as we did for the quasiparticle hierarchy with non-abelian condensation as we did in the previous section. With these choices of $m$ and $l$, the fusion product of $\sigma_2\v a^{-1}\sigma_1'\v b^l$ with itself does not have any bosons. It follows from (\ref{ninequ}) that this does not give a valid condensation. 

Like in the previous section, we can try to construct a condensable algebra by stacking with a FQH state based on the $SU(3)_2$ MTC. We find that for $\sigma_2\v a^{-1}\mathbf{8}\v b^l$ to be a charge zero boson,
\begin{equation}
    m=5(10n-7)v^2\qquad l=(10n-7)v.
\end{equation}

This leads to hierarchy states at filling fraction
\begin{equation}
    \nu_n=\frac{6n-4}{10n-7},
\end{equation}
which match with the filling fractions in (\ref{z3ab}). However, these states differ from those in (\ref{z3ab}) because they are abelian. For $n=1$ and $v=2$, we have $m=60$ and $l=6$. We propose the condensable algebra in $SU(3)_{2}\times U(1)_{60}$ stacked with the minimal modular extension of the $\mathbb{Z}_3$ Read-Rezayi state:
\begin{equation}
    \mathcal{A}=\mathbf{0}+\psi_1\v a^{-1}\v b^6+\sigma_2\v a^{-1}\mathbf{8}\v b^6+\psi_2\v a^{-2}\v b^{12}+\sigma_1\v a^{-2}\mathbf{8}\v b^{12}+\cdots.
\end{equation}

The resulting theory is abelian, with chiral central charge $c_-=6$, and is generated by $\mathbf{6}$ and $\v b^{10}$. The latter has the anyon content of $U(1)_{-3}$ after condensing the fermion $\v b^{30}$.

If instead we condense $\sigma_2\v a^{-1}\mathbf{3}\v b^l$, then we would get
\begin{equation}
    m=\frac{5}{3}(30n-11)v^2\qquad l=\frac{1}{3}(30n-11)v.
\end{equation}

This gives states at filling fraction
\begin{equation}
    \nu_n=3\left(\frac{6n-2}{30n-11}\right),
\end{equation}
which are less likely to be observed compared to those above.
\section{Beyond minimal quasihole/quasiparticle: Bonderson-Slingerland hierarchy}\label{sbondsling}
Previously, we always considered condensing the minimal quasihole or quasiparticle. In this section, we will consider some examples involving condensing non-minimal quasiholes and quasiparticles. This reproduces the hierarchy constructed in Ref.~\cite{bonderson2008} for the Pfaffian/anti-Pfaffian states. Unlike the hierarchy states in Sec.~\ref{spfaff}, these hierarchy states can be non-abelian. This construction was used to explain the non-abelian $\nu=7/3$ and $\nu=8/3$ states with minimal quasihole charge of $\frac{1}{3}$, and also to explain the $\nu=2+3/8$ plateau.

In Sec.~\ref{spfaff}, we considered condensing the minimal quasihole $\sigma \v a$. To reproduce the results of Ref.~\cite{bonderson2008}, we will instead condense the quasihole $\v a^2$, which is not minimal, and carries charge $\frac{1}{2}$. For the quasihole hierarchy, we stack with $U(1)_{-m}$. For $\v a^2\bar{\v b}^l$ to be a charge zero boson, we get
\begin{equation}\label{bshole}
m=2(4n+1)v^2\to\nu_n=\frac{2n}{4n+1}.
\end{equation}

This matches the quasihole hierarchy of Ref.~\cite{levin2009}. The $n=1$ hierarchy state has filling fraction $2+\frac{2}{5}$. If we choose $\v v=\v b$, then $m=10$ and $l=5$, so we condense $\v a^2\bar{\v b}^{5}$. However this anyon fuses with itself to give $\v a^4$, which has odd integer charge and cannot be condensed. Therefore, we instead choose $\v v=\v b^2$, giving $m=40$ and $l=10$. After condensing the condensable algebra generated by $\v a^2\bar{\v b}^{10}$, and the resulting topological order is non-abelian. $\sigma \v a\v b$ braids trivially with all the condensed anyons, and has charge $\frac{1}{4}-\frac{1}{20}=\frac{1}{5}$, in agreement with Ref.~\cite{bonderson2008}.

For the quasiparticle hierarchy, we stack with $U(1)_m$ and find
\begin{equation}
m=2(4n-1)v^2\to\nu_n=\frac{2n}{4n-1}.
\end{equation}

Let us consider the $n=1$ hierarchy state, with filling fraction $2+\frac{2}{3}=\frac{8}{3}$. Again, we must choose $\v v=\v b^2$, to obtain a condensable algebra with only anyons carrying zero charge. $\v v=\v b^2$ means $m=24$ and we condense $\v a^2\v b^{-6}$. The non-abelian anyon $\sigma \v a\v b$ braids trivially with this anyon, so it survives the condensation. This anyon has charge $\frac{1}{4}+\frac{1}{6}=\frac{1}{3}$, consistent with Ref.~\cite{bonderson2008}. 

The $\nu=2+\frac{3}{8}$ state appears as a second level hierarchy. Specifically, starting from the $\frac{2}{5}$ (non-minimal) quasihole hierarchy state, we condense a minimal quasihole. The $\frac{2}{5}$ state has a minimal quasihole $\bar{\v b}^{-4}$ (where $\bar{\v b}$ is the generator of $U(1)_{-m}$ and $m$ is written in (\ref{bshole})), which has charge $\frac{1}{5}$. Since it is abelian, we stack with $U(1)_{-m}$, generated by $\bar{\v c}$. We find that condensing $\bar{\v b}^{-4}\bar{\v c}^l$ requires
\begin{equation}
    m=10(5n-1)v^2\qquad l=2(5n-1)v.
\end{equation}

For $n=1$, this gives a filling fraction $\frac{2}{5}+\frac{1}{40}=\frac{3}{8}$.
 
\section{Flux attachment via anyon condensation}\label{sflux}
In this section, we describe a way to apply flux attachment from an algebraic perpsective. Flux attachment maps between a fermionic wavefunction with $p$ Landau levels of composite fermions (where the composite fermions are formed by attaching $2q$ fluxes to each electron) to a fractional quantum Hall state at filling fraction $\nu=\frac{p}{2pq+1}$ by
\begin{equation}
    \Psi_{p/(2pq+1)}(\{z_i\})=\mathcal{P}_{\mathrm{LLL}}\left[\chi_1^{2q}(\{z_i\})\chi_p(\{z_i\})\right].
\end{equation} 

It is a map that preserves chiral central charge (for the positive Jain sequence at $\frac{p}{2pq+1}$), but changes filling fraction, and changes the object identified with the physical fermion. These properties indicate that the flux attachment procedure should involve stacking with a nonchiral theory and condensing a condensable algebra that preserves the new physical fermion. We will expand on this idea in this section. We will first show how flux attachment works for $p=1$, and then we will generalize to other $p$.

To understand what flux attachment means in the algebraic description of the FQH phase, we need to first describe integer quantum Hall states. The $\nu=1$ integer quantum Hall state is described by the sMTC $\{1,f\}$ where $f$ is a transparent fermion and the vison. In order to remove $f$ to obtain a different transparent fermion, we need to introduce anyons that braid nontrivially with $f$. In other words, we need a minimal modular extension of $\{1,f\}$. We choose the minimal modular extension $U(1)_4$ because it has a chiral central charge matching the free fermion $\nu=1$ state. The theory has anyons $\{\v a^i\}$ for $i=0,1,2,3$, and the fermion is identified with $i=2$. $i=1$ and $i=3$ correspond to fermion parity fluxes, and braid with $f=\v a^2$ with a phase of $-1$. 

We now need to attach fluxes to the fermion $f=\v a^2$. We stack with $U(1)_{8q}$, generated by $\v b$, and choose the new electron to be $f \v b^{4q}$. Since $\v b^{4q}$ is a boson, $f\v b^{4q}$ also has fermionic statistics. In order for this fermion to carry charge 1, we choose the vison to be $\v b^2$. This symmetry fractionalization pattern means that every insertion of a $2\pi$ flux corresponds to an insertion of $\v b^2$, so choosing the new transparent fermion to be $ f\v b^{4q}$ means that we attach $2q$ fluxes to the original electron.

We can force $f\v b^{4q}$ to be the new physical electron by condensing $\v a\v b$, which braids nontrivially with $f$ but trivially with $f\v b^{4q}$. In order to make this a boson (and maintain $c_-=1$), we stack with $U(1)_{-m}$, generated by $\bar{\v c}$. In order for $\v a\v b\bar{\v c}^l$ to be a charge zero boson, we find
\begin{equation}
    m=2q(16qn+2q+1)v^2\qquad l=\frac{1}{2}(16qn+2q+1)v,
\end{equation}
where $n$ is an integer. For the smallest positive $m$, we choose $n=0$ and $v=2$. This results in $m=8q(2q+1)$ and $l=2q+1$. The filling fraction is
\begin{equation}
    \nu=\frac{1}{2q}-\frac{1}{2q(2q+1)}=\frac{1}{2q+1},
\end{equation}
which is the Jain sequence for $p=1$ (where the composite fermions completely fill a single Landau level). Projecting out anyons that braid nontrivially with the new fermion $f\v a^4$ produces the fermionic theories corresponding to the Jain sequence.

To access other integer $p$, we can stack $U(1)_4$ theories to obtain integer quantum Hall states with higher Hall conductance. Every time we stack with an additional $U(1)_4$ factor, to increase $p$ by one, we must condense the fermion-fermion pair $\v a_1^2\v a_2^2$ in order to have a single fermion. Then repeating the above process for $p$ layers of $U(1)_4$ theories gives
\begin{equation}
    \nu=\frac{p}{2pq+1},
\end{equation}
which match the positive Jain sequence.

The flux attachment procedure is powerful because relative to the hierarchy constructions presented earlier in this work, it preserves the stability of the states well. Specifically, the gap of a Jain sequence state can be related to the gap of the corresponding integer quantum Hall state it is derived from, labeled by $p$, as has been confirmed in experiment\cite{du1993}. Therefore, although the algebraic theory of anyons is generally agnostic to energetics, this particular algebraic manipulation seems to preserve stability better than the hierarchy constructions studied in the bulk of this work, at least according to experimental and numerical results.

Notice that this is consistent with the observation that the Pfaffian quasihole hierarchy state has a Jain sequence filling fraction. To produce the Pfaffian quasihole hierarchy states, we took $\mathrm{Ising}\times U(1)_8\times\overline{\mathrm{Ising}}\times U(1)_{-m}$. In Sec.~\ref{spfaff}, we wrote a condensable algebra including $\sigma$ and $\bar{\sigma}$, but as we discussed there, we can also obtain the hierarchy states by sequential abelian condensation. First condensing $\psi\bar{\psi}$ produces $\mathbb{Z}_2$ gauge theory stacked with $U(1)_8\times U(1)_{-m}$. The $\mathbb{Z}_2$ gauge theory corresponds to eight copies of $U(1)_4$ (with the appropriate fermion-fermion condensations) as we described above. The hierarchy construction and the flux attachment construction both involve condensing the same subgroup in this abelian theory, so they result in the same anyon theory but with $c_-$ differing by eight. Specifically, the difference in chiral central charge by eight comes from the fact that in the Pfaffian hierarchy construction we view $\mathbb{Z}_2$ gauge theory has having $c_-=0$ while in the flux attachment picture we view $\mathbb{Z}_2$ gauge theory as coming from eight integer quantum Hall systems with $c_-=8$. A similar argument explains why the other hierarchy states obtained by replacing $\mathrm{Ising}$ by other theories in Kitaev's 16-fold way\cite{kitaev2006} produce hierarchy states that match with Jain filling fractions\cite{zhelt2024}. In Appendix~\ref{s16fold}, we reproduce these hierarchies by stacking the FQH states based on the even elements of the 16-fold way with $U(1)_{\pm m}$ and by stacking the FQH phases based on the odd elements of the 16-fold way  with $\mathrm{Ising}\times U(1)_m$ and $\overline{\mathrm{Ising}}\times U(1)_{-m}$. However, it should be emphasized that the Jain sequence is built on integer quantum Hall states labeled by integers, not (half) integers modulo eight. 

\section{Discussion}\label{sdiscussion}
We have presented a method for constructing FQH hierarchies by splitting the "anyon condensation into FQH phase" process into two well-defined steps. The first step involves stacking the parent FQH state with a second FQH state, such that the total filling is the sum of the original filling and the filling of the new state. The second step is to condense a condensable algebra of charge-neutral bosons, resulting in a simpler state at the same filling. While the bulk of this paper focused on the case where the condensable algebra is generated by an anyon that is a fusion product between the minimal quasihole or quasiparticle in the parent state and an anyon in the stacked theory, we also discuss some generalizations beyond this scenario in Sec.~\ref{sbondsling}. This approach allows for the filling fractions and topological data of the hierarchy states to be calculated more easily, without the need to manipulate explicit wavefunctions. One benefit of working without a wavefunction is that we can apply our method to the PH-Pfaffian state. Many of the filling fractions we obtain for hierarchy states match with filling fractions of other known states, for example at Jain state fillings. While at the same fillings, some of these states are quite different, and even might be non-abelian. There are also examples where the anyon data precisely match with a Jain state, but the chiral central charge differs by a eight (such as for the Pfaffian hierarchies).

While computationally easier, the approach is more abstract, so it is also harder to determine which mathematically sound predictions are realistically favored predictions. One direction for future work is to check the overlap of the hierarchy states presented here with numerical ground states. The above hierarchy states tend to have rather large rank. One can test the hierarchy construction by numerically by studying the states near the $\mathbb{Z}_3$ Read-Rezayi state numerically. It would be especially interesting to test if any of the filling fracion $\frac{1}{3}$ or $\frac{2}{3}$ have non-abelian theories matching those presented in this work. Since the hierarchy states for the fermionic Pfaffian state have rather high rank, one can instead try to numerically study the states near the bosonic Pfaffian. The bosonic Pfaffian has filling fraction $\nu=1$, and has smaller rank than the fermionic Pfaffian. The hierarchy states also have smaller rank, occuring at $\nu=\frac{8}{9}$ and $\nu=\frac{7}{6}$ (see Sec.~\ref{sbosonicpfaff}); these states might be amenable to a more accurate numerical study.

In this work, we also presented, to our knowledge, the first examples of hierarchy states from minimal quasihole/quasiparticle condensation that are non-abelian. For example, we obtained non-abelian hierarchies for the $\mathbb{Z}_3$ Read-Rezayi state and the FQH state based on the $SU(2)_3$ CFT. It would be especially interesting to try to search for these states numerically or experimentally. In both of the FQH states mentioned above, there are multiple minimal quasiholes/quasiparticles, so there are multiple options for condensable algebras. It would be interesting to better understand if certain condensations (i.e. the abelian condensations, which lead to non-abelian hierarchy states) are favored over others. 

This work used the more formal tools developed in the mathematical literature to reinterpret the hierarchy construction and other wavefunction manipulations such as flux attachment. We hope that this opens the door to the exploration of other applications of these formal tools. On the other hand, the physics of the hierarchy construction also motivates the following question: can we generalize the mathematics of non-abelian anyon condensation to include the condensation of anyons with nontrivial spin, in the presence of a magnetic field? In other words, can we combine the two-step procedure described here into a more general, unified picture of anyon condensation? Similarly, the physics of the flux attachment procedure motivates the question: can we better understand what kinds of algebraic manipulations are better at preserving stability in the sense described at the end of Sec.~\ref{sflux}? More generally, it would be very interesting to develop an understanding of wavefunction manipulations generalizing flux attachment, such as multiplying more general wavefunctions and projecting into a Landau level. These procedures can be used to obtain many non-abelian states; it would be useful to understand what they mean from an algebraic perspective. 

There are also other quantities of interest in the FQH literature. In particular, the shift of a FQH state describes how rotation symmetry is fractionalized in the system\cite{wen1992}. While there is a systematic theory of shift for bosonic systems\cite{manjunath2020} which involves identifying a particular abelian anyon $s$ that assigns fractional orbital angular momentum quantum numbers, there is not yet one for fermionic systems. Interestingly, shift in fermionic systems generally cannot be described by identifying a particular abelian anyon $s$ inserted at $2\pi$ curvature flux. While the shift can be read off from a concrete wavefunction, in an algebraic description of a FQH phase, it is an extra piece of data that must be specified. It would be interesting to generalize the analysis of this paper to include shift. For example, one might have additional constraints on possible anyon condensation processes coming from a zero (fractional) orbital angular momentum condition.

Finally, we comment on the physical interpretation of our hierarchy construction. On the one hand, the theories $\eD$ that we stack with tend to be quite complicated and high rank (such as $\overline{\mathrm{Ising}}\times U(1)_{-136}$), so one might be tempted to view $\eD$ as just a computational tool. On the other hand, the important role of these high rank topological orders in determining hierarchies may imply that such theories should be taken more seriously, and might appear under appropriate conditions. It would be especially interesting to see if there are signatures of the stacked theory $\eD$ and the condensation of $\mathcal{A}$ at the transition between a FQH state and its neighboring hierarchy state, either numerically or experimentally.

\section*{Acknowledgements}

We thank B. Halperin and N. Manjunath for helpful discussions, and D. Son for suggesting studying the bosonic Pfaffian state.
C.Z. is supported by the Harvard Society of Fellows, and thanks the Perimeter Institute for hosting the workshop ``Higher Categorical Tools for Quantum Phases of Matter,'' where participants provided valuable feedback. This work is partially supported by NSF DMR-2022428 and by the Simons 
Collaboration on Ultra-Quantum Matter, which is a grant from the Simons 
Foundation (651446, XGW). We also acknowledge support from a Simons Investigator grant (AV) and from NSF-DMR 2220703.

\appendix
\section{Wavefunction derivation}\label{swavefunction}

In many cases, the FQH wavefunction can be obtained using the CFT ansatz \cite{moore1991,hansson2017}. In this section, we will focus on the FQH states where this is possible. For these states, we will obtain the hierarchy construction, including the filling fractions of the hierarchy states, at the level of the wavefunction. 

The basic idea is as follows. We first consider the wavefunction of the parent
state $\eC$. In this parent state there will be a particular
anyon $\v a$ that we want to condense. To derive hierarchy states, we will choose $\v
a$ to be the anyon with smallest positive or negative charge in $\eC$. To condense this anyon, we
introduce a second theory $\eD$ with an anyon $\v b$ such that $\v a \v b$ is a
boson. A more subtle requirement is that $\v a \v b$ must generate a
condensable algebra in $\eC\boxtimes \eD$.\footnote{Note that requiring $\v a\v b$ to generate a condensable algebra is a stronger constraint than having $\v a\v b$ belong in a condensable algebra.}\footnote{For a discussion about how a boson may not give rise to a
condensable algebra, see Sec.~\ref{sconstraints}}  Even if $\v a \v b$ is a boson, it cannot
condense if it sees a magnetic field. The requirement that $\v a \v b$ sees
zero average magnetic field gives a constraint relating the number of $\v a\v
b$ bosons, which we denote by $N_b$, and the number of electrons $N_e$.
Intuitively, the density of $\v a\v b$ gives an effective magnetic field when
$\v a\v b$ is pulled around the system, that cancels with the enclosed flux.
This constraint allows us to solve for the total angular momentum of the
theory, which gives the filling fraction of the daughter states. These filling fractions match with those described in the main text, and we will relate these two ways of deriving filling fraction at the end of this appendix.

We now work out the details of the above physical picture. Consider a parent
non-abelian FQH state of charge 1 (we set $e=1$) electrons described by wavefunction
$\Psi(z_1,\cdots,z_{N_e})$, which has a form
\begin{align}
 \Psi(z_1,\cdots,z_{N_e}) = 
 P(z_1,\cdots,z_{N_e}) e^{-\frac14 \sum |z_i|^2}.
\end{align}

The Gaussian factor $ \ee^{-\frac14 \sum |z_i|^2}$ will not be important in the following discussion. We will focus on the properties of $P(z_1,\cdots,z_{N_\text{e}})$, which is an analytic function of the electron positions $\{z_i\}$. We will take as an ansatz for this function the correlation function of primary fields in a CFT\cite{wen1994}:
\begin{align}\label{electronwf}
 P(z_1,\cdots,z_{N_e}) = \lim_{z_\infty \to \infty}
\left\<  e^{-N_\text{ele}\ii \frac{\phi(z_\infty)}{\sqrt{\nu}}}
\prod_{i=1}^{N_e} V_e(z_i)
\right\>.
\end{align}

There is a close relation between 1+1D CFTs and 2+1D topological orders: primary fields in the CFT
correspond to 2+1D anyons. Fermionic systems form supermodular tensor categories (sMTC)\cite{moore1989,witten1989}, while bosonic systems form modular tensor categories (MTC). Therefore, we can write down the wavefunction of a FQH state with certain anyons by using correlation functions in a CFT with the same modular data. 

In (\ref{electronwf}), $V_e(z_i)$ is the primary field for electrons. It has the form 
\begin{align}
 V_e(z) = \psi(z) e^{\ii \frac{ \phi(z)}{\sqrt{\nu}} },
\end{align}
where $\psi$ is a simple current operator in the CFT.\footnote{Simple currents
are operators with pointed (abelian) fusion.} $\phi$ is the scalar field in a
free boson CFT.  There is always a free boson CFT, like how there is always an abelian $U(1)$ sector of the anyon theory for FQH phases. Here, we normalize $\phi$ such that $e^{\ii \phi}$ has a
scaling dimension $\frac12$, \ie
\begin{align}
\lim_{z_1\to z_2} \left\< e^{\ii \phi(z_1)} e^{-\ii \phi(z_2)} \right\> \sim \frac{1}{z_1-z_2}.
\end{align}

For a given operator $\psi(z_i)$, $\nu$ must be chosen such that $ V_e(z_i) $
and $V_e(z_j)$ are mutually local. In other words, $ P(z_1,\cdots,z_{N_e})$
must be a single-valued function of the electron positions $\{z_i\}$. The chiral boson
part contributes a factor of $\prod_{i<j}(z_i-z_j)^{1/\nu}$ to
$P(z_1,\cdots,z_{N_e})$, so the total power of $z_i$'s is proportional to
$N_e^2$ to leading order.  Specifically, the total power of $z_i$'s is $\frac
{N_e^2}{2} \nu^{-1} +O(N_e)$.  Therefore, the filling fraction of the
non-abelian state is given by $\nu$ as in the Laughlin state. Note that the
operators $\psi(z_i)$ can only modify the total power of $z_i$'s by an amount
of order $N_e$.)

Now that we have completed our description of the FQH wavefunction, we can
choose excitations $\v a \in \eC$ of the parent FQH state to condense.  In
the CFT description of FQH state, such an excitation is described a primary field
\begin{align}
\label{Va}
 V_{\v a}(\xi) = \psi_{\v a} (\xi) e^{-\ii q \frac{\phi(\xi)}{ \sqrt{\nu}} },
\end{align}
where $\psi_{\v a}$ is a primary field (not necessarily simple) in the CFT. From the form of the operator above, we see that an anyon $\v a$
carries an electric charge $q$. $q>0$ correspond to a quasiparticle and
$q<0$ correspond to a quasihole. 

The many-electron wavefunction with excitations at coordinates $\{\xi_i\}$ is given by
\begin{align}
&\ \ \ \
 P(z_1,\cdots,z_{N_e}; \xi_1,\cdots,\xi_{N_{\v a}}) 
\nonumber\\
&
= 
\lim_{z_\infty \to \infty}
\<  \ee^{-(N_e-N_{\v a} q )\ii \frac{\phi(z_\infty)}{ \sqrt{\nu}}}
\prod_{i=1}^{N_e} V_e(z_i)
\prod_{j=1}^{N_{\v a}} V_{\v a}(\xi_i)
\>.
\end{align}

We need to choose $\psi_{\v a}$ and $q$ such that $ V_{\v a}(\xi)$ and $V_e(z)$
are mutually local, meaning $ P(z_1,\cdots,z_{N_\text{ele}};
\xi_1,\cdots,\xi_{N_{\v a}})$ is a single-valued function of $z_i$'s.  Note
that  $ P(z_1,\cdots,z_{N_e}; \xi_1,\cdots,\xi_{N_{\v a}})$ may not be a
single-valued function of the $\xi_i$'s. 

To condense the $\v a$ anyons, we introducing a new CFT $\eD$ and consider a
primary field $V_{\v b}$ in $\eD$.  We choose $V_{\v b}$ such that $V_{\v
a}V_{\v b}$ is bosonic (or more precisely, we require that the corresponding
anyons $\v a\otimes \v b$ generates a condensible algebra in $\eC\boxtimes
\eD$). We then use correlation function
\begin{align}
\label{Pztz}
\<  
\prod_{i=1}^{N_e} V_e(z_i)
\prod_{j=1}^{N_{b}} V_{\v a}(\t z_i)V_{\v b}(\t z_i)
\>
\end{align}
to construct the wave function $ P(z_1,\cdots,z_{N_e}; \t z_1,\cdots,\t
z_{N_{b}})$ for the daughter states of the parent FQH state, induced by the
condensation of the algebra generated by $\v a\otimes \v b$.

To be more concrete, let us first assume $\v a$-anyon to be a quasi-particle
and consider quasi-particle condensation. In other words, we assume $q>0$.  

Let us also assume that the primary field $V_{\v b} \in \eD$ has the form
\begin{align}\label{vbform}
 V_{\v b}(\t z) = \t \psi_{\v b}(\t z) 
\ee^{\ii \t q \t \phi(\t z)}.
\end{align}
We like to choose $ V_{\v b}$ such that $V_{\v a} V_{\v b}$ is bosonic. More
precisely, we require that there is a $n$ such that\footnote{This is related to the $n$-cluster condition discussed in Refs.~\cite{wen2008,wen2008zeros}.}
\begin{align}
 (V_{\v a} V_{\v b})^n \sim 
\ee^{- n \ii  q \frac{\phi(\tilde{z})}{ \sqrt{\nu}} }
\ee^{n \ii \t q \t \phi(\t z)} + \cdots.
\end{align}

Inserting these $\v a\v b$ operators gives a FQH state with the wavefunction given in \eq{Pztz}. Let us focus on the chiral boson part of the wavefunction, to compute the filling fractions of the hierarchy states. The chiral boson part is given by
\begin{align}
\label{ztzxi}
&
\Psi(\{z_i\},\{\tilde{z}_i\})\propto  \prod_{1\leq i < j \leq N_e} (z_i-z_j)^{\nu^{-1}}
\\
& \ \ \ \ 
 \prod_{1\leq i < j \leq N_{b}} (\t z_i-\t z_j)^{\frac{q^2}{\nu} +\t q^2}
\hskip -3mm
\prod_{1\leq i \leq N_e,1\leq j\leq N_{b}} 
( z_i-\t z_j)^{-q \nu^{-1} } .
\nonumber 
 \end{align}

Note that we omitted the contributions to the wavefunction from $\psi$, $\psi_{\v a}$, and $\t\psi_{\v b}$, because they do not contribute to the chiral boson part. The first term in (\ref{ztzxi}) describes correlations between electrons, the second term describes correlations between quasiparticles, and the third term describes correlations between electrons and quasiparticles.

To condense the anyons $\v a \v b$, we must integrate over the quasiparticle positions $\{\tilde{z}_i\}$. This results in the daughter state wavefunction, which is a wavefunction that only depends on
the electron coordinates $\{z_i\}$.

To integrate over $\{\t z_i\}$, we require that the $\v a\v b$ anyons see zero average magnetic field, so that modifying $\{\t z_i\}$ does not give the wavefunction too many phase oscillations.  To enforce this condition, we assume
that the $N_e$ electrons and $N_{b}$ anyons form two disks of the same size.  Moving an anyon around the edge of the disk should not change the phase of the wavefunction. This condition gives
\begin{align}
\left(\frac{q^2}{\nu}+\t q^2\right) N_b - \frac{q}{ \nu} N_e = 0,
\end{align}
where the first term is the braiding phase between the anyon along the edge and those in the bulk of the disk and the second term is the braiding phase from the anyon on the edge and the electrons. Since $q >0$, the above equations have a solution with $N_e, N_b >0$:
\begin{align}\label{NbNe}
N_b &=   \frac{q}{q^2+\t q^2 \nu} N_e.
\end{align}

The total angular momentum of the wavefunction (\ref{ztzxi}) is, at leading order,
\begin{align}
 L_\text{tot} &= 
\frac1{2\nu} N_e^2 
+ \frac{\frac{q^2}{\nu}+\t q^2}{2} N_{b}^2 
- \frac{q}{\nu} N_{b} N_e 
\nonumber \\
& =
\frac12 N_e^2 \left[]
\frac1{\nu} + \frac{(\frac{q^2}{\nu}+ \t q^2) q^2}{(q^2+\t q^2\nu)^2 }
- 2\frac{q}{\nu} \frac{q}{q^2+\t q^2 \nu}
\right]
\nonumber\\
&=
\frac12 N_e^2 \left(
\frac1{\nu} - \frac{q^2}{q^2\nu+\t q^2 \nu^2}
\right)
.
\end{align}
Now using $\nu=\frac{1}{2}N_e^2/L_{\mathrm{tot}}$, we see that the filling
fraction of the quasi-particle condensation-induced daughter state is
\begin{align}\label{quasiparticlenu}
 \nu_\text{tot} = \frac{1}{\frac1{\nu} - \frac{q^2/\nu^2}{\frac{q^2}{\nu}+\t q^2 }}.
\end{align}

Suppose that $V_{\v b}(\tilde{z})$ has charge of the same magnitude $q$ as $V_{\v a}(z)$. This makes $\v a\v b$ charge neutral.
In this case, $\t q$ in \eqn{vbform} has a form
\begin{equation}
\tilde{q}=\frac{q}{\sqrt{\tilde{\nu}}},
\end{equation}
where $\tilde{\nu}$ has a meaning of the filling fraction of the theory $\eD$.
Plugging this into (\ref{quasiparticlenu}) gives the simplified expression
\begin{align}
\nu_\text{qp} &= \frac{1}{\frac1{\nu} 
- \frac{q^2/\nu^2}{\frac{q^2}{ \nu}+\frac{q^2}{\tilde{\nu}} }} 
= \frac{1}{\frac1{\nu} - \frac{1}{ \nu+\frac{\nu^2}{\tilde{\nu}}}} 
=\nu+\tilde{\nu}.
\end{align}

Next, we consider quasi-hole condensation, i.e. we assume $q<0$.  In this case, we
need to choose $\eD$ to be an anti-holomorphic CFT.  The primary field $V_{\v
b}$ now has a form
\begin{align}\label{fieldanti}
 V_{\v b}(\t z^*) = \t \psi_{\v b}(\t z^*) \ee^{\ii \t q \t \phi(\t z^*)}.
\end{align}

The wavefunction with anyon insertions now has the following $U(1)$ part:
\begin{align}\label{u1parthole}
& \prod_{1\leq i < j \leq N_e} (z_i-z_j)^{\nu^{-1}}
 \prod_{1\leq i < j \leq N_b} 
(\t z_i-\t z_j)^{\frac{q^2}{\nu}}
(\t z_i^*-\t z_j^*)^{\t q^2}
\nonumber\\
&
\prod_{1\leq i \leq N_e,1\leq j\leq N_b} 
( z_i-\t z_j)^{-q \nu^{-1} }.
 \end{align}
 
Again, the first term comes from electron-electron correlations, the second term comes from anyon-anyon correlations, and the third term comes from correlations between anyons and electrons.

Like in the quasiparticle case, we need to make sure moving the anyon positions do not give the wavefunction too many phase oscillations. This means that we need the following condition to be satisfied:
\begin{align}
\left(\frac{q^2}{\nu}-\t q^2\right) N_{ b} - \frac{q}{\nu} N_e = 0,
\end{align}
where again the first term comes from the braiding phase between an anyon along the edge of the disk and anyons in the bulk and the second comes from the braiding phase between an anyon along the edge and electrons in the bulk. The above equations have the solution solution 
\begin{align}\label{nbneconstr}
  N_{b} &=  \frac{q}{q^2-\t q^2\nu}  N_e.
\end{align}

The total angular momentum of the condensation-induced wave function is again given by counting powers in (\ref{u1parthole}) with the constraint (\ref{nbneconstr}):
\begin{align}
 L_\text{tot} &= 
\frac1{2\nu} N_e^2 
+ \frac{\frac{q^2}{\nu}-\t q^2}{2} N_{b}^2 
- \frac{q}{\nu} N_{ b} N_e
\nonumber \\
& =
\frac12 N_e^2 \left[
\frac1{\nu} + \frac{(\frac{q^2}{\nu}- \t q^2) q^2}{(q^2-\t q^2\nu)^2 }
- 2\frac{q}{\nu} \frac{q}{q^2-\t q^2 \nu}
\right]
\nonumber\\
&=
\frac12 N_e^2 \left(
\frac1{\nu} - \frac{q^2}{q^2\nu-\t q^2 \nu^2}
\right)
.
\end{align}

We see that the filling fraction of the quasihole condensation-induced
daughter state is
\begin{align}\label{quasiholenu}
 \nu_\text{qh} = \frac{1}{\frac1{\nu} - \frac{q^2/\nu^2}{\frac{q^2}{\nu}-\t q^2}}.
\end{align}

Suppose that we are considering quasihole condensation, so $V_{\v
b}(\tilde{z})$ has opposite charge compared to $V_{\v a}(z)$, so that again $\v a\v b$ is charge neutral. Then in (\ref{fieldanti}), we have
\begin{equation}
\tilde{q}=\frac{q}{\sqrt{\tilde{\nu}}},
\end{equation}
where $-\tilde{\nu}$ is the filling fraction of the theory $\eD$. In this case, (\ref{quasiholenu}) simplifies to
\begin{align}
 \nu_\text{qh} &= \frac{1}{\frac1{\nu} - \frac{q^2/\nu^2}{\frac{q^2}{\nu}-\frac{q^2}{\tilde{\nu}}}}
 = \frac{1}{\frac1{\nu} - \frac{1/\nu^2}{\frac{1}{\nu}-\frac{1}{\tilde{\nu}}}}
=\nu+(-\tilde{\nu}).
\end{align}

\subsubsection{Wavefunction derivation of hierarchy states for abelian FQH phases}
We can also derive the results in Sec.~\ref{slaughlin} at the level of the wavefunction, following Appendix~\ref{swavefunction}. The primary field for electrons is given by
\begin{align}
 V_e(z) = \ee^{\ii \sqrt p_1 \phi(z)}.
\end{align}

It follows that the wavefunction of the $\nu=\frac{1}{p_1}$ Laughlin state (neglecting the Gaussian part) is given by
\begin{align}
 P(z_1,\cdots,z_{N_e}) = \prod_{i<j} (z_i-z_j)^{p_1} .
\end{align}

The condensing anyon $\v a$ has minimal negative charge $-\frac{1}{p_1}$, and is given by
primary field
\begin{align}
 V_{\v a}(z) = \ee^{-\ii \frac{\phi(z) }{\sqrt{p_1}} }.
\end{align}

We also choose the $V_{\v b}$ to be given by primary field
\begin{align}
 V_{\v b}(\t z) = \ee^{\ii \t q\t\phi(\t z)}.
\end{align}

In order for $V_{\v a} V_{\v b}$ to be bosonic,
we require
\begin{align}
\frac{ \t q^2}{2} =  -\frac1{2p_1}  + n ,
\ \ \to \ \
\t q^2 = -\frac1p_1+2n ,\ \ 
. 
\end{align}

Using (\ref{quasiparticlenu}), we find the filling fraction of the daughter state of the quasiparticle condensation to be
\begin{align}
 \nu_{\mathrm{tot}} 
= \frac{1}{p_1 - \frac{1}{2n} }
= \frac{2k}{2np_1-1}.
\end{align}

This recovers the results in (\ref{slaughlin}) for the quasiparticle hierarchy.

Next we choose the condensing anyon $\v a$ to be a quasi-hole given by
primary field
\begin{align}
 V_{\v a}(z) = \ee^{\ii \frac{\phi(z) }{\sqrt p_1} },
\end{align}
\ie we choose $q=-\frac{1}{p_1}$.
We also choose the $V_{\v b}$ to be given by primary field
\begin{align}
 V_{\v b}(\t z^*) = \ee^{\ii \t q\t\phi(\t z^*)}.
\end{align}

In order for $V_{\v a} V_{\v b}$ to be bosonic,
we require
\begin{align}
\frac{ \t q^2}{2} = \frac1{2p_1}   +n,
\ \ \to \ \
\t q^2 = \frac1p_1 +2n,
. 
\end{align}

Using (\ref{quasiholenu}), we compute the filling fraction of the daughter state of the quasihole condensation to be
\begin{align}
 \nu_{\mathrm{tot}} 
= \frac{1}{p_1 + \frac{1}{2n} }
= \frac{2n}{2np_1+1} .
\end{align}

Further details about the resulting theory can also be rederived from the wavefunction perspective.
\section{$S$ and $T$ matrices}\label{sstmat}
In this appendix, we review a necessary but not sufficient condition for a condensable algebra $\mathcal{A}$ to describe a gapped boundary between $\eC$ and $\eC/\mathcal{A}$, based on $S$ and $T$ matrices. More details can be found in Appendix C of Ref.~\cite{ng2023}. We will not have to check this condition for the particular condensations described in this paper, because we use only condensations that are combinations of diagonal condensations (of the form $\sum a\bar{a}$ where $\bar{a}$ is the time reversal of $a$ and we sum over all of the simple objects of an MTC) and (sequences of) abelian condensations. Diagonal condensable are well-known to satisfy all the criteria for gapped domain walls.

Let us denote the MTC obtained from condensing $\mathcal{A}$ in $\eC$ by $\eC/\mathcal{A}$. Condensable algebras describing domain walls between $\eC$ and $\eC/\mathcal{A}$ correspond to gapped boundaries of the folded theory $\eC\boxtimes\overline{\eC/\mathcal{A}}$. The gapped boundaries of this theory are labeled by integer vectors with $\mathrm{rank}(\eC)\times\mathrm{rank}(\eC/\mathcal{A})$ entries:
\begin{equation}\label{algstack}
    \mathcal{A}_{\eC\boxtimes\overline{\eC/\mathcal{A}}}=\oplus_{\v a\in\eC,\v y\in\overline{\eC/\mathcal{A}}}\v A^{\v a\v y}\v a\otimes\v y.
\end{equation}

A necessary (but not sufficient) condition for $\v A$ to describe a gapped boundary of $\eC\boxtimes\overline{\eC/\mathcal{A}}$ is
\begin{equation}\label{smatcond}
    S_{\eC\boxtimes\overline{\eC/\mathcal{A}}}\v A=\v A\qquad T_{\eC\boxtimes\overline{\eC/\mathcal{A}}}\v A=\v A.
\end{equation}

Using the fact that the $S$ matrix for two stacked theories is simply the tensor product of the $S$ matrix for each individual theory (and the $T$ matrix has the same property), we can write (\ref{smatcond}) as 
a constraint on $\v A$ reshaped into a $\mathrm{rank}(\eC)\times\mathrm{rank}(\eC/\mathcal{A})$ matrix:
\begin{equation}\label{scaint}
S_{\eC}\v A=\v A S_{\eC/\mathcal{A}},
\end{equation}
\begin{equation}\label{tconstraint}
T_{\eC}\v A=\v A T_{\eC/\mathcal{A}},
\end{equation}
where we use a slight abuse of notation: in $\v A$ of the two equations above, we take $\v y\to\bar{\v y}\in\eC/\mathcal{A}$ compared to (\ref{algstack}). We will explain the conditions (\ref{scaint}) and (\ref{tconstraint}) in more detail in Appendix~\ref{sstmat}. One immediate consequence of (\ref{scaint}) that is worth pointing out here. (\ref{scaint}) can be written as
\begin{equation}\label{smatind}
    \sum_{\v b\in\eC}S_{\eC}^{\v a\v b}\v A^{\v b\v y}=\sum_{\v z\in\eC/\mathcal{A}}\v A^{\v a\v z}S_{\eC/\mathcal{A}}^{\v z\v y}.
\end{equation}

As discussed in Appendix~\ref{sstmat}, there is a particularly simple canonical domain wall between $\eC$ and $\eC/\mathcal{A}$, describing the setup where all excitations in $\eC/\mathcal{A}$ can pass into $\eC$ without leaving any nontrivial excitations at the boundary. This canonical domain wall is given by
\begin{equation}\label{vadef}
\v A^{\v a\v 0}=n_{\v a},
\end{equation}
and for any $\v y\in\eC/\mathcal{A}$, there exists a $\v a\in\eC$ such that $\v A^{\v a\v y}\neq 0$. The $\v a$ for which $\v A^{\v a\v y}\neq 0$ are the anyons that tunnel through the domain wall from $\v a\in\eC$ to $\v y\in\eC/\mathcal{A}$. In other words, they are the anyons that do not get confined by the condensation.

For the canonical domain wall, setting $\v a=\v y=\v 0$ in (\ref{smatind}) gives (\ref{dimchange}).

\section{Condensable algebras in fermionic topological orders}\label{sfermionic}
In this appendix, we describe two approaches to formalizing condensable algebras for fermionic theories. We mostly use approach 1 in the paper. Approach 1 is based on minimal modular extensions, which are bosonic theories obtained by gauging the fermion parity of a fermionic theory. Approach 2 works directly in the fermionic theory, presenting constraints on the (degenerate) $S$ and $T$ matrices of the fermionic theory.
\subsection{Approach 1: minimal modular extension}\label{smme}
Our first approach to using the machinery of condensable algebra in sMTCs is to first extend the sMTC $\eC$ to a modular theory $\eC^{(b)}$. This is known as minimal modular extension. Every sMTC has 16 minimal modular extensions, corresponding to gauging fermion parity in the theory stacked with different numbers of $p_x+ip_y$ superconductors. For example, the trivial fermionic theory $\eC=\{\v 0,\v f\}$ has 16 minimal modular extensions given by Kitaev's 16-fold way\cite{kitaev2006}. Modular extensions are bosonic (braiding nondegenerate) theories that contain the anyons of the fermionic theory:
\begin{equation}
\eC\subset\eC^{(b)},
\end{equation}
and minimal modular extensions are $\{\eC^{(b)}\}$ with the smallest rank.

After obtaining a minimal modular extension $\eC^{(b)}$, we can stack with a bosonic $\eD$ and condense a bosonic condensable algebra in $\eC^{(b)}\boxtimes\eD$. This produces a bosonic daughter state. We can then obtain the corresponding fermionic one by condensing the fermion, removing anyons that braid nontrivially with the electron operator.

In summary, this procedure involves the following steps:
\begin{enumerate}
\item Find a minimal modular extension $\eC^{(b)}$ of $\eC$
\item Stack $\eC^{(b)}$ with a MTC $\eD$
\item Condense $\mathcal{A}$ in $\eC^{(b)}\boxtimes\eD$
\item Condense the fermion to obtain a fermionic theory
\end{enumerate}

In general there is no direct recipe for finding the minimal modular extensions, given a sMTC. However, the FQH systems of interest here are all either abelian, or of the form $\eC\boxtimes U(1)_m/\mathbb{Z}_2$ where $\eC$ and $U(1)_m$ are both MTCs. The mod $\mathbb{Z}_2$ gives the fermion condensation, removing anyons that braid nontrivially with the electron operator. Therefore, the simplest minimal modular extension in this case is obtained by removing the mod $\mathbb{Z}_2$, to recover the original bosonic theory. 

One caveat with using this approach is that one must check that the hierarchies do not depend on the choice of minimal modular extension. We ensure this in our hierarchies because the anyons that we condense all belong in the original fermionic theories; none of them include anyons added by minimal modular extension.

\subsection{Approach 2: fermionic condensable algebras}\label{sfermioniccond}
In this section we will work directly with fermionic theories and adapt (\ref{scaint}) and (\ref{tconstraint}) to sMTCs. In a sMTC, the $S$ and $T$ matrices take the form
\begin{equation}
S=\frac{1}{\sqrt{2}}\begin{pmatrix} 1 & 1\\ 1 & 1\end{pmatrix}\otimes \hat{S}\qquad T=\begin{pmatrix} 1 & 0\\ 0 & -1\end{pmatrix}\otimes\hat{T},
\end{equation}
where $\hat{S}$ and $\hat{T}$ are unitary. Note that given $T$, $\hat{T}$ is not well defined. However, $\hat{T}^2$ is well defined.  

In sMTCs, we can add fermions to $\mathcal{A}$, as long as they satisfy (\ref{rconstraint}). We will identify anyons that differ by the transparent fermion, so the number of distinct anyon types is give by the rank of $\hat{S}$ rather than the rank of $S$. We claim that the following condition must be satisfied as the fermionic analogue of (\ref{scaint}) and (\ref{tconstraint}):
\begin{equation}\label{sfermionic}
\hat{S}_{\eC\boxtimes\eD}\mathbf{A}=\mathbf{A}\hat{S}_{\eC\boxtimes\eD/\mathcal{A}},
\end{equation}
\begin{equation}\label{tfermionic}
\hat{T}_{\eC\boxtimes\eD}^2\mathbf{A}=\mathbf{A}\hat{T}_{\eC\boxtimes\eD/\mathcal{A}}^2.
\end{equation}

Note that (\ref{tfermionic}) naturally allows $\mathcal{A}$ to include fermionic anyons (although we do not do so here because they carry odd integer charge). 

(\ref{sfermionic}) and (\ref{tfermionic}) are necessary conditions for $\mathcal{A}$ to extend into a condensable algebra of $\eC^{(b)}$ that includes only the subset of anyons in $\eC$. This is clear from the observation that a modular extension only adds anyons to $\eC$ without changing the fusion and braiding properties of the anyons originally in $\eC$. Therefore, the $S$ and $T$ matrices of a modular extension always has a block that is identical to the $S$ and $T$ matrices of $\eC$. This fact has been used to derive modular extensions from the modular data of the sMTC. 

In more detail, a minimal modular extension adds fermion parity fluxes given by $v$-type anyons (which fuse with the fermion to a distinct flux) and $\sigma$-type anyons (which absorb the fermion). It was shown in Ref.~\cite{seo2024} that $S_{\eC^{(b)}}$ and $T_{\eC^{(b)}}$ take the form
\begin{equation}
S_{\eC^{(b)}}=\begin{pmatrix} \frac{1}{2}\hat{S} & \frac{1}{2}\hat{S} & A & A & X\\ \frac{1}{2}\hat{S} & \frac{1}{2}\hat{S} & -A & -A & -X\\ A^T & -A^T & B & -B & 0\\ A^T & -A^T & -B & B & 0\\ X^T & -X^T & 0 & 0 & 0\end{pmatrix},
\end{equation}
\begin{equation}
T_{\eC^{(b)}}=\mathrm{diag}\begin{pmatrix}\hat{T} & -\hat{T} & T_v & T_v & T_{\sigma}\end{pmatrix}.
\end{equation}

The particular form of the matrices $A,X,B,T_v,$ and $T_{\sigma}$ will not be important for our purposes, because our condensable algebras only include anyons in $\eC$. In $\eC^{(b)}$, we have a new, larger matrix $\v A^{(b)}$, which is just the original matrix (\ref{vadef}) except with additional zero matrix elements. From the form of the matrices $S_{\eC^{(b)}}$ and $T_{\eC^{(b)}}$ above, it is clear that $\v A$ consisting only of bosons, satisfying (\ref{sfermionic}) and (\ref{tfermionic}), also satisfies (when extended with zero matrix elements) (\ref{scaint}) and (\ref{tconstraint}) with $S_{\eC^{(b)}}$ and $T_{\eC^{(b)}}$.

\section{Constructing the $GL(N,\mathbb{Z})$ basis change matrix}\label{sgln}
In this appendix we explain how to derive integer basis change matrices like (\ref{xmat}) to map between the $K$ matrices of theories with anyon condensation to the $K$ matrices of their daughter states.

Specifically, we would like to find a $GL(N,\mathbb{Z})$ matrix $X$ that maps
\begin{equation}
\tilde{K}_H=\begin{pmatrix} \tilde{K} & l\\ l^T & 0\end{pmatrix}\to K_{\mathrm{qp}}=\begin{pmatrix} 0 & \v 0^T\\ \v 0^T & K_{\mathrm{qp}}'\end{pmatrix},
\end{equation}
where $K_{\mathrm{qp}}'$ is a $(N-1)\times (N-1)$ matrix consisting of a daughter state block $\begin{pmatrix} p_1 & -1\\ -1 & p_2\end{pmatrix}$ together with trivial blocks. Here, $l$ is an integer vector describing the anyon that generates the condensable subgroup $\mathcal{A}$. It has zero spin and charge.

Without loss of generality, we can write each column of $X$ as $c_i=\begin{pmatrix} v_i' & n_i\end{pmatrix}^T$, where $v_i'$ is an integer vector of length $N-1$ and $n$ is an integer. We can always write $v=\tilde{K}v'\to v'=\tilde{K}^{-1} v$ where $v$ is a trivial anyon in the theory described by $\tilde{K}$. Then setting $X^T\tilde{K}_H X=K_{\mathrm{qp}}$, we obtain the following matrix elements of $K_{\mathrm{qp}}$:
\begin{align}
\begin{split}\label{matelement}
\left(K_{\mathrm{qp}}\right)_{ij}&=c_i^T\tilde{K}_H c_j\\
&=v_i^T\tilde{K}^{-1}v_j+n_jv_i^T\tilde{K}^{-1}l+n_il^T\tilde{K}^{-1}v_j.
\end{split}
\end{align}

We see that this is just the braiding phase of an anyon labeled by $c_i$ with an anyon labeled by $c_j$, as long as $l$ has spin zero (not just integer): $l^T K^{-1}l=0$. The anyon $c_i$ is given by the trivial anyon $v_i$ fused with $n_i$  copies of $l$, and the anyon $c_j$ is given by the trivial anyon $v_j$ fused with $n_j$ copies of $l$. Note that if $l$ did not have spin zero, then the above braiding phase interpretation of (\ref{matelement}) would be off by $n_in_jl^T\tilde{K}^{-1}l$. 

The fact that the top row and first column of $K_{\mathrm{qp}}$ consists entirely of zeros means that the first column of $X$, $c_0$, must be the vacuum. We can choose $v_0=\mathrm{ord}(l)l\to v_0'=\mathrm{ord}(l)\tilde{K}^{-1}l$ and $n_0=-\mathrm{ord}(l)$, where $\mathrm{ord}(l)$ is the order of the condensed anyon $l$.

The other columns $c_i$ can be obtained by solving (\ref{matelement}) with the desired $K_{\mathrm{qp}}$. For the example in (\ref{xmat}), we need $c_1$ to be an anyon with spin $e^{2\pi i 3/2}$, $c_2$ to be an anyon with spin $e^{2\pi i}$, and for them to have mutual braiding $e^{-2\pi i}$. We can choose $c_1$ to be $a^{-3}$, which has spin $e^{2\pi i 9/6}=e^{2\pi i 3/2}$ and we can choose $c_2$ to be $ab^{-5}$ which has spin $e^{2\pi i/6}e^{2\pi i 25/30}=e^{2\pi i}$. $a^{-3}$ is already a trivial anyon. We can also write $ab^{-5}$ as trivial anyons together with copies of the condensed one $ab^{-5}cd^{-1}$ (where $c$ and $d$ are excitations of the two negative integer quantum Hall states): $ab^{-5}=(ab^{-5}cd^{-1})c^{-1}d$. This gives $v_1=\begin{pmatrix}-3 & 0 & 0 & 0 & 0 & 0\end{pmatrix}^T$ and $v_2=\begin{pmatrix}0 & 0 & -1 & 1 & 0 & 0\end{pmatrix}^T$, giving $c_1=\begin{pmatrix}-1 & 0 & 0 & 0 & 0 & 0 & 0 \end{pmatrix}^T$ and $c_2=\begin{pmatrix}0 & 0 & 1 & -1 & 0 & 0 & 1\end{pmatrix}^T$. We find that $c_3,c_4,c_5,$ and $c_6$ can be identified each of the four integer quantum Hall states, because they all have spin $e^{\pm 2\pi i/2}$ and trivial mutual braiding. This gives (\ref{xmat}).

%

\section{16-fold way}\label{s16fold}
In Sec~\ref{spfaff} and Sec.~\ref{ssu22}, we focused on FQH theories based on $\mu=1$ and $\mu=3$ in Kitaev's 16-fold way\cite{kitaev2006} (here we use $\mu$ instead of the usual symbol $\nu$ because we already use $\nu$ to refer to the filling fraction) together with an abelian $U(1)_8$ theory. In this section, we will first reproduce the results from Ref.~\cite{zhelt2024}, which considered all values of $\mu\in[0,16)$. As we will show below, the filling fractions for the hierarchy states for different values of $\mu$ match the Jain sequence $\frac{p}{2pq\pm 1}$ with $q$ equal to 1. In the Jain sequence, $q$ refers to the number of vortices or antivortices attached to each electron, and $p$ refers to the number of filled composite fermion Landau levels. We will then show that generalizing $U(1)_8$ to $U(1)_{8q}$ produces hierarchy states at filling fraction $\frac{p}{2pq+1}$ for all integer $q$. 

The even values of $\mu$ label abelian topological orders, while the odd ones label non-abelian topological orders. For the abelian ones, we can reproduce the hierarchies from Ref.~\cite{zhelt2024} by stacking with $U(1)_{\pm m}$ while for the non-abelian ones, we can reproduce the hierarchies from Ref.~\cite{zhelt2024} by stacking with $\mathrm{Ising}\times U(1)_{m}$ and $\overline{\mathrm{Ising}}\times U(1)_{-m}$. We stack with these theories because they lead to the smallest change in chiral central charge. Replacing $\mathrm{Ising}$ by another non-abelian theory in the 16-fold way would cause a larger jump in chiral central charge between the parent state and its hierarchy states.

The hierarchy states have the same filling fraction as the Jain sequence, but generally differ from the Jain states by chiral central charge that is a multiple of 8, i.e. an integer number of E8 states. From the wavefunction calculation in Ref.~\cite{levin2009}, they also differ in their shift\cite{levin2009}, which is related to fractionalization of spatial symmetries. It follows that stacking one of these hierarchy states (i.e. that of the Pfaffian state) with a Jain state at the charge conjugate filling, and then condensing pairs of anyons between the two layers, may produce an E8 state, in the absence of additional spatial symmetries.

\subsection{$q=1$ Jain fillings}\label{sq1}
We begin with the abelian theories, where $\mu$ is an even integer. These theories have abelian anyons $\{\mathbf{0},\psi,\v e,\v m\}$ with spin $\{1,-1,e^{\frac{2\pi i\mu}{16}},e^{\frac{2\pi i\mu}{16}}\}$. When $\mu=0$, the theory is simply the toric code, or $\mathbb{Z}_2$ gauge theory. Like for the Pfaffian state, we add a $U(1)_{8}$ theory generated by the anyon $\v a $, with the vison chosen to be $\v a^2$ to obtain $\nu=\frac{1}{2}$. To obtain a fermionic theory, we also project out anyons that braid nontrivially with the physical fermion $\psi \v a^4$. The minimal quasiholes after projection are $\v e\v a$ and $\v m\v a$ while the minimal quasiparticles are $\v e\v a^{-1}$ and $\v m\v a^{-1}$. To construct the hierarchies, we stack with $U(1)_{\pm m}$, generated by $\v b$. Requiring $\v e\v a\v b^l$ to be a charge zero boson gives the solutions
\begin{equation}\label{mueqjain}
    m=2(16n\mp(\mu+1))v^2\qquad l=\frac{1}{2}(16n\mp(\mu+1))v,
\end{equation}
where $(-)$ gives the quasiparticle hierarchy and $(+)$ gives the quasihole hierarchy. It follows that the filling fractions of these hierarchy states are
\begin{equation}
    \nu_n=\frac{1}{2}\left(\frac{16n\mp \mu}{16n\mp(\mu+1)}\right),
\end{equation}
matching the results from Ref.~\cite{zhelt2024}. Writing $\mu=2\tilde{\mu}$ gives, for the quasihole hierarchy,
\begin{equation}\label{qhjain}
    \nu_n=\frac{(8n+\tilde{\mu})}{2(8n+\tilde{\mu})+1},
\end{equation}
with $c_-=\tilde{\mu}$. These states have the same filling fraction as the Jain sequence, with $8n+\tilde{\mu}$ identified with the number of filled composite fermion Landau levels (with the composite fermions consisting of electrons each attached to two vortices). The chiral central charge of the Jain state at filling $\frac{(8n+\tilde{\mu})}{2(8n+\tilde{\mu})+1}$ is $8n+\tilde{\mu}$, so the Jain states at the same filling as these hierarchy states have an extra chiral central charge of $8n$ (i.e. $n$ copies of $E8$ states, in the absence of other symmetries and superconductivity). For $n=0$, the above states are precisely the Jain states with $\tilde{\mu}=p,q=1$.

For the quasiparticle hierarchy, (\ref{mueqjain}) simplifies to
\begin{equation}\label{qpjain}
    \nu_n=\frac{(8n-\tilde{\mu})}{2(8n-\tilde{\mu})-1},
\end{equation}
with $c_-=\tilde{\mu}+2$. The Jain state with filling fraction $\frac{8n-\tilde{\mu}}{2(8n-\tilde{\mu})-1}$ has chiral central charge $\tilde{\mu}+2 - 8 n $, so the Jain states differ from those in (\ref{qpjain}) in that they have an extra $n$ antichiral E8 states. The precise Jain states correspond to the particle-hole conjugate of the states described above.

Odd values of $\nu$ correspond to non-abelian theories in the 16-fold way\cite{kitaev2006}. These theories have three anyons $\{\mathbf{0},\psi,\sigma\}$ with spins $\{1,-1,e^{\frac{2\pi i\mu}{16}}\}$. When $\mu=1$, the theory is Ising topological order. Like for the Pfaffian state, we add another $U(1)_8$ theory generated by $\v a$ with the vison chosen to be $\v a^2$, and we project out anyons that braid nontrivially with the physical fermion $\psi \v a^4$. We can obtain a quasihole hierarchy by stacking with $\overline{\mathrm{Ising}}\times U(1)_{-m}$. This gives
\begin{equation}
    m=2(16n+\mu)v^2\qquad l=\frac{1}{2}(16n+\mu)v,
\end{equation}
so we have
\begin{equation}\label{non-abqhjain}
    \nu_n=\frac{1}{2}\left(\frac{16n+ \mu-1}{16n+\mu}\right).
\end{equation}

For $\mu=1$, we recover the Pfaffian quasihole hierarchy. In terms of $\mu-1=2\tilde{\mu}$, (\ref{non-abqhjain}) simplifies to
\begin{equation}
    \nu_n=\frac{8n+\tilde{\mu}}{2(8n+\tilde{\mu})+1},
\end{equation}
with chiral central charge $c_-=\tilde{\mu}$. The Jain states at the same filling differ from these states in that they have an extra chiral central charge of $8n$. For the quasiparticle hierarchy, we stack with $\mathrm{Ising}\times U(1)_m$. The solutions are given by 
\begin{equation}
    m=2(16n-\mu-2)v^2\qquad l=\frac{1}{2}(16n-\mu-2)v.
\end{equation}

This gives states at filling fraction 
\begin{equation}
    \nu_n=\frac{8n-\tilde{\mu}-1}{2(8n-\tilde{\mu}-1)-1},
\end{equation}
with $c_-=\tilde{\mu}+3$. The Jain states at the same filling have an extra $-8n$ chiral central charge. 

\subsection{$q>1$ Jain fillings}\label{sjaingeneral}
In the previous section, we obtained the Jain sequence $\frac{p}{2pq+1}$ with $q=1$ (with the possible addition of E8 states) by considering hierarchy states of FQH theories built out of a topological order in the 16-fold way together with a $U(1)_8$ abelian topological order. In this section, we will show that using $U(1)_{8q}$ instead of $U(1)_8$ leads to hierarchy states with the topological order at more general Jain states, with  filling fraction $\frac{p}{2pq+1}$ (again, with the chiral central charge that generally differs by eight times an integer).

If we replace $U(1)_8$ by $U(1)_{8q}$, we can construct a fermionic FQH theory by identifying the physical fermion with $\psi \v a^{4q}$ and projecting out anyons that braid nontrivially with this fermion. In order for the fermion to have odd integer charge, we choose the vison to be $\v a^2$. Since the spin of the vison is $\frac{1}{4q}$, this symmetry fractionalization pattern gives a theory with filling fraction $\frac{1}{2q}$. Like in the previous section, we can take $\psi$ from any of the 16 theories in the 16-fold way. Different choices of $\mu\in[0,16)$ and $q\in\mathbb{Z}^+$ produce the Jain filling fractions $\frac{p}{2pq\pm 1}$. 

For $\mu$ even, the minimal quasiholes after projection are again $\v e\v a$ and $\v m\v a$. We stack with $U(1)_{\pm m}$ generated by $\v b$. In order for $\v e\v a\v b^l$ to be a charge zero boson, we must have
\begin{align}
\begin{split}    
m&=2q(16nq\mp(q\mu+1))v^2\\
l&=\frac{1}{2}(16nq\mp(q\mu+1))v.
\end{split}
\end{align}

The hierarchy states have filling fraction
\begin{equation}
\nu_n=\frac{(8n\mp\tilde{\mu})}{2(8n\mp\tilde{\mu})q\mp1},
\end{equation}
where $\tilde{\mu}=\mu/2$. 

For $\mu$ odd, we again stack with $\overline{\mathrm{Ising}}\times U(1)_{-m}$ for the quasihole hierarchy. This gives solutions
\begin{align}
\begin{split}
m&=2q(16nq+\mu q+1-q)v^2\\
l&=\frac{1}{2}(16nq+\mu q+1-q)v.
\end{split}
\end{align}

The quasihole hierarchy states occur at filling fractions
\begin{equation}
    \nu_n=\frac{8n+\tilde{\mu}}{2(8n+\tilde{\mu})q+1},
\end{equation}
where $\tilde{\mu}=\frac{1}{2}(\mu-1)$. By a similar calculation, the quasiparticle hierarchy states occur at filling fractions
\begin{equation}
    \nu_n=\frac{8n-\tilde{\mu}-1}{2(8n-\tilde{\mu}-1)q-1}.
\end{equation}

Therefore, the filling fractions of the hierarchy states match with those of the Jain sequence with $q\geq 1$. However, since the chiral central charges of these theories generally differ by eight times an integer.

\section{Derivation of the $SU(2)_3$ state}\label{ssu23cft}

The $SU(2)_3$ CFT has primary fields $\{\mathbf{0},\mathbf{\frac{1}{2}},\mathbf{1},\mathbf{\frac{3}{2}}\}$ with scaling dimensions $\{0,\frac{3}{20},\frac{2}{5},\frac{3}{4}\}$.
To get the electron operator, we choose a combination of $\mathbf{\frac{3}{2}}$ and a chiral bose vertex $V_{\alpha}$:
\begin{equation}
\psi_e(z)=\mathbf{\frac{3}{2}}(z)V_{\alpha}(z).
\end{equation}

The OPE gives
\begin{equation}
\psi_e(z)\psi_e(w)\sim (z-w)^{-3/2}(z-w)^{\alpha^2}V_{2\alpha}.
\end{equation}

For this to be single valued, we need $\alpha^2=3/2+n$ where $n$ is an integer. The theory is fermionic if $n$ is an even integer as discussed in Ref.~\cite{blok1992} (naively, one would think $n$ should be odd to get an antisymmetric wavefunction, but as explained in Ref.~\cite{blok1992}, we must choose $n$ to be even if we include the fact that the electrons carry a spin $\frac{3}{2}$ representation of $SU(2)$). For $n=0$, we have
\begin{equation}
\psi_e(z)=\mathbf{\frac{3}{2}}(z)V_{\sqrt{3/2}}(z).
\end{equation}

Because the $SU(2)_3$ fields and the bose fields are completely separate, the wavefunction for $N$ electrons is then given by
\begin{equation}
\Psi=\left\langle\mathbf{\frac{3}{2}}(z_1)\cdots\mathbf{\frac{3}{2}}(z_N)\right\rangle\prod_{i<j}(z_i-z_j)^{3/2}\prod_{i=1}^Ne^{-|z_i|^2/(4l^2)}.
\end{equation}

The filling fraction can be read off as $\nu=\frac{2}{3}$ as expected. To get the charge of the minimal quasihole, we note that $\mathbf{\frac{1}{2}}$ braids with phase -1 with $\mathbf{\frac{3}{2}}$ so we attach it with a primary field as an ansatz for the quasihole:
\begin{equation}
\psi_{qh}(w)=\mathbf{\frac{1}{2}}(w)V_{\beta}(w)
\end{equation}
The OPE then gives
\begin{equation}
\psi_{qh}(w)\psi_e(z)\sim(w-z)^{-1/2}(w-z)^{\beta\sqrt{3/2}}
\end{equation}
In order for this wavefunction to not have any branch cuts for the physical electron, we must have $\beta=(n+1/2)/\sqrt{\frac{3}{2}}$ where the minimal quasihole is given by $n=0$. We find that the minimal quasihole has charge $e*(2/3)/2=e/3$. The primary field $V_{1/2\sqrt{3/2}}=V_{1/\sqrt{6}}$ indicates that we will attach the $SU(2)_3$ theory with $U(1)_6$ (with a transparent fermion $f$). We obtain the theory with anyons listed in (\ref{su23anyons}), where $\bar{\mathbf{\frac{3}{2}}}\bar{a}^3$ is the transparent fermion.

\bibliography{hierarchy.bib}

\end{document}

%% file: defs.tex
\usepackage{graphicx}
\usepackage{dcolumn}
\usepackage{bm}
\usepackage{multirow}

\usepackage[normalem]{ulem}

\usepackage{amsmath}
\usepackage{amsthm}
\usepackage{amstext}
\usepackage{amssymb}
\usepackage{mathrsfs}
\usepackage{amsfonts}
\usepackage{amsbsy} 

\usepackage[all,matrix,cmtip]{xy}

\usepackage{csquotes}
\MakeOuterQuote{"}

\usepackage{color}
\usepackage[colorlinks,citecolor=blue,linkcolor=blue,urlcolor=blue]{hyperref}

\definecolor{red}{rgb}{1,0,0}
\definecolor{blue}{rgb}{0,0,1}
\definecolor{dblue}{rgb}{0,0,0.4}
\definecolor{green}{rgb}{0,1,0}
\definecolor{dgreen}{rgb}{0,0.4,0}
\definecolor{black}{rgb}{0,0,0}
\definecolor{white}{rgb}{1,1,1}

\definecolor{brn}{rgb}{.8,.4,.0}
\definecolor{redo}{rgb}{1,.5,.0}
\definecolor{ddgrn}{rgb}{0,0.4,0}
\definecolor{dgrn}{rgb}{0,0.55,0}
\definecolor{dbl}{rgb}{0,0,0.5}
\definecolor{grey}{rgb}{0.5,0.5,0.5}

\usepackage[bbgreekl]{mathbbol}

\renewcommand{\v}[1]{\boldsymbol{#1}} 
\renewcommand{\t}[1]{\widetilde{#1}} 
\newcommand{\ii}{\hspace{1pt}\mathrm{i}\hspace{1pt}}
\newcommand{\ee}{\hspace{1pt}\mathrm{e}}

\newcommand{\<}{\langle} 
\renewcommand{\>}{\rangle} 

\newcommand{\Rf}[1]{Ref.~\onlinecite{#1}}
 
\newcommand{\eq}[1]{(\ref{#1})} 
\newcommand{\eqn}[1]{eqn.~(\ref{#1})}

\newcommand{\ie}{{\it i.e.~}}

\newcommand{\bpm}{\begin{pmatrix}}
\newcommand{\epm}{\end{pmatrix}}
\newcommand{\bmm}{\begin{matrix}}
\newcommand{\emm}{\end{matrix}}

\newcommand{\cA}{ {\cal A} }

\usepackage{euscript}

\newcommand\eC           {\EuScript{C}}
\newcommand\eD           {\EuScript{D}}